\documentclass[12pt,a4paper]{report}

\usepackage[top=0.70in, bottom=0.70in, left=0.8in,right=0.80in]{geometry} 
\usepackage[pdftex]{graphicx} 
\usepackage[final]{pdfpages} 
\usepackage{float} 
\usepackage{pslatex} 
\usepackage{array} 
\usepackage{setspace}
\usepackage{float}
\usepackage{enumerate}
\usepackage{longtable}
\newcommand{\eml}{\textsc{Embedded Matlab}}
\newcommand{\prover}{\textsc{Prover}}

\newcommand{\ccsl}{\textsc{Ccsl}}
\newcommand{\tdl}{\textsc{Tadl}}

\newcommand{\xtc}{\textsc{Xtc}}
\newcommand{\gt}[1]{\texttt{#1}}
\newcommand{\s}{\textsc{S/S}}
\newcommand{\fp}{\textsc{$f_p$}}
\newcommand{\av}{\textsc{AV}}

\newcommand{\uppaal}{\textsc{Uppaal}}
\newcommand{\ode}{\textsc{ODE}}
\newcommand{\smc}{\textsc{Uppaal-smc}}

\newcommand{\simu}{\textsc{Simulink}}
\newcommand{\staf}{\textsc{Stateflow}}
\newcommand{\sldv}{\textsc{Simulink  Design Verifier}}
\newcommand{\sdv}{\textsc{SDV}}
\newcommand{\me}{\textsc{Metaedit+}}
\newcommand{\mt}{\textsc{Matlab}}
\newcommand{\ed}{\textsc{East-adl}}

\newcommand{\ft}{\textsc{$F_t$}}

\usepackage{makeidx}  
\usepackage{url}
\usepackage{graphicx}
\usepackage{subfigure}
\usepackage{makecell}
\usepackage{cite}
\usepackage{listings}
\usepackage{paralist}
\let\subparagraph\paragraph
\usepackage{cite}
\usepackage{amsfonts, amsmath, amssymb}
\usepackage{mathptmx}
\usepackage{algorithm}
\usepackage{algorithmic}
\usepackage{epstopdf}
\usepackage{float}
\usepackage{url}
\usepackage{enumerate}
\usepackage{color}
\usepackage{multirow}
\usepackage{verbatim}
\usepackage{bm}
\usepackage{longtable}
\usepackage[figuresright]{rotating}
\usepackage{geometry}
\usepackage{authblk}

\usepackage[font=small,labelfont=bf]{caption}

\usepackage{listings}
\usepackage{color}

\definecolor{dkgreen}{rgb}{0,0.6,0}
\definecolor{gray}{rgb}{0.5,0.5,0.5}
\definecolor{mauve}{rgb}{0.58,0,0.82}

\usepackage{fancyhdr}
\fancypagestyle{plain}{%
\fancyfoot[L]{\emph{}} 
\fancyfoot[R]{\thepage}

}

\usepackage{pgf}
\usepackage{pgfpages}

\pgfpagesdeclarelayout{boxed}
{
  \edef\pgfpageoptionborder{0pt}
}
{
  \pgfpagesphysicalpageoptions
  {%
    logical pages=1,%
  }
  \pgfpageslogicalpageoptions{1}
  {
    border code=\pgfsetlinewidth{2pt}\pgfstroke,%
    border shrink=\pgfpageoptionborder,%
    resized width=.95\pgfphysicalwidth,%
    resized height=.95\pgfphysicalheight,%
    center=\pgfpoint{.5\pgfphysicalwidth}{.5\pgfphysicalheight}%
  }%
}
\begin{document}
\renewcommand\bibname{References}
\lhead{ }



\newpage
\begin{center}
\thispagestyle{empty}
\setlength{\voffset}{2in}
\vspace{2cm}
\LARGE{\textbf{Model-based Verification and Validation of an Autonomous
Vehicle System: Simulation and Statistical Model Checking}}\\
\vspace{2cm}
{\Large{\textbf{Eun-Young Kang$^{12}$, Dongrui Mu$^{2}$, Li Huang$^{2}$ and Qianqing Lan$^{2}$}}}\\
\vspace{0.3cm}
\large{\textbf{$^1$PReCISE Research Centre,
University of Namur, Belgium}}\\
\vspace{0.3cm}
\large{\textbf{$^2$School of Data and Computer Science, \\Sun Yat-sen University, Guangzhou, China}}\\
\vspace{0.3cm}
\Large{{eykang@fundp.ac.be}}\\
\Large{{\{mudr, huangl223, lanqq\}@mail2.sysu.edu.cn}}\\
\vspace{3cm}
\newpage

\end{center}
\newpage



\begin{center}
\thispagestyle{empty}
\vspace{2cm}
\LARGE{\textbf{ABSTRACT}}\\[1.0cm]
\end{center}
\thispagestyle{empty}
\large{\paragraph{}
The software development for Cyber-Physical Systems (CPS), e.g., autonomous vehicles, requires both functional and non-functional quality assurance to guarantee that the CPS operates safely and effectively.  \ed\ is a domain specific architectural language dedicated to safety-critical automotive embedded system design. We have previously modified \ed\ to include energy constraints and transformed energy-aware real-time (ERT) behaviors modeled in \ed/\staf \\ into \uppaal~models amenable to formal verification. Previous work is extended in this paper by including support for \simu\ and an integration of Simulink/Stateflow within a same tool-chain. Simulink/Stateflow models are transformed, based on extended ERT constraints in \ed, into verifiable \uppaal~models with stochastic semantics and integrate the translation with formal statistical analysis techniques: Probabilistic extension of \ed\ constraints is defined as a semantics denotation.  A set of mapping rules is proposed to facilitate the guarantee of translation. Formal analysis on both functional- and non-functional properties is performed using \sldv/\smc. The analysis techniques are validated and demonstrated on the autonomous traffic sign recognition vehicle case study.}

\textbf{Keywords: }CPS, \ed, \smc, \sldv, Verification \& Validation
\newpage

\pagenumbering{roman} 

\pagestyle{empty}
\addtocontents{toc}{\protect\thispagestyle{empty}}
\tableofcontents 

\addtocontents{lof}{\protect\thispagestyle{empty}}
\listoffigures 
\cleardoublepage

\pagestyle{fancy}

\newpage
\pagenumbering{arabic} 

\chapter{Introduction}
The software development for Cyber-Physical Systems (CPS) requires both functional and non-functional quality assurance to guarantee that CPS operates in a safety-critical context under timing and resource constraints. Automotive electric/electronic systems are ideal examples of CPS, in which the software controllers interact with physical environments. The continuous time behaviors (evolved with various energy/cost rates) of those systems often rely on complex dynamics as well as on stochastic behaviors. Formal verification and validation (V\&V) technologies are indispensable and highly recommended for development of safe and reliable automotive systems \cite{iso26262,iec61508}. Conventional V\&V have limitations in terms of assessing the reliability of hybrid systems due to the both stochastic and non-linear dynamical features. To assure the reliability of safety critical hybrid dynamic systems, \emph{statistical model checking (SMC)} techniques have been proposed \cite{smc-david-12,david2015uppaal}. These techniques for fully stochastic models validate probabilistic performance properties of given deterministic (or stochastic) controllers in given stochastic environments.

\ed\ \cite{EAST-ADL} is a domain specific language
that provides support for architectural specification and timing behavior constraints for automotive embedded systems. A system in \ed\ is
described by {\gt{Functional Architectures (FA)}} at different
abstraction levels. The {\gt{FAs}} are composed of a number of
interconnected \emph{functionprototypes} (\fp), and
the \fp s have ports and connectors for communication.
\ed\ relies on external
tools for the analysis of specifications related to requirements or safety, V\&V. For example,
behavioral description in \ed\ is captured in external tools, i.e., Simulink/Stateflow \cite{slsf}. The latest release of \ed\ has adopted the time model proposed in the Timing Augmented Description Language (\tdl2) \cite{TADL2}. \tdl2 allows for the expression and composition of timing constraints, e.g., repetition rates, end-to-end delay, and synchronization constraints on top of \ed\ models.

Our previous work in \cite{kqsic12,ksac14,kapsec15,kiciea16} extended \ed\ timed behavior constraints with energy consumption constraints and
introduced transformation algorithms which map energy-aware real-time (ERT) behaviors (modeled in UML/Stateflow)  to the \uppaal\ \cite{uppaal} models. The results are used as the basis of the current approach, which include support for \simu\ and an integration of Simulink/Stateflow (S/S) within a same tool-chain. S/S models are transformed, based on the extended ERT constraints with probability parameters, into verifiable \smc\ \cite{smc} and integrate the translation with formal statistical analysis techniques: Probabilistic extension of \ed\ constraints is defined as a semantics denotation.  A set of mapping rules is proposed to facilitate the guarantee of translation. Formal analysis on both functional- and non-functional properties is performed using \sldv\ (\sdv) \cite{SLDV} and \smc. Our approach is demonstrated on the autonomous traffic sign recognition vehicle (\av) case study and identifies potential conflicts between different automotive functions.

The paper is organized as follows: Chapter \ref{sec:background}
presents an overview of \simu\ and \smc. The \av\ case study is
introduced as a running example in Chapter \ref{sec:case-study}.
Chapter \ref{sec:ss-trans} and \ref{sec:smc-trans} describe a set of mapping patterns and how our modeling approach provides
support for simulation and formal analysis at the design level. The applicability of our method is demonstrated by performing V\&V on the \av\ case study in Chapter \ref{sec:v-v}. We discuss related work in Chapter \ref{sec:r-work}. The conclusion is presented in Chapter \ref{sec:conclusion}.

\chapter{Background}
\label{sec:background}

In our framework, \simu\ and \eml\ are utilized for modeling purposes. Simulation and V\&V are performed by \simu\ and \smc.

\section{\simu/\eml}

\simu\ is a synchronous data flow language, which provides different types of \emph{blocks}, i.e., primitive-,
control flow-, and temporal-blocks with predefined libraries for modeling and simulation of dynamic systems and
code generation. \simu\ supports the definition of custom blocks via \staf\ diagrams or
\emph{user-defined function} blocks, namely \emph{S-Functions}, written in \eml\ (EML), C, C++, or
Fortran. Each primitive block has a corresponding construct in EML except the \emph{unit delay} block, which delays
 and holds its input signal by one sampling interval. Sample time \cite{sampletime} is a parameter of a block
 indicating when, during simulation, the block generates outputs and if appropriate, updates its internal state
 during simulation. The dynamic models can be simulated and the results displayed as simulation runs.

\section{\sldv}

\sdv\ is a plug-in to \prover, which is a formal verification tool that performs reachability analysis. The satisfiability of each reachable state is determined by a SAT solver. \sdv\ generates tests for \simu/\eml\ models according to model coverage and user-defined objectives.

\begin{figure}[hbp]
\centerline{{\includegraphics[width=5in]{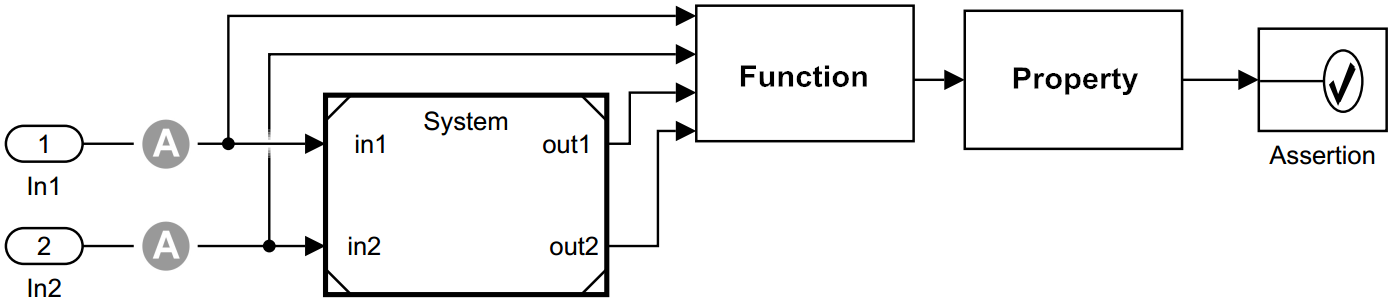}}}
\caption{Proof outline in Simulink/SDV}
\label{fig_general_verification_model}
\end{figure}

A proof objective is specified in \simu/\sdv\ and illustrated in Fig.\ref{fig_general_verification_model}. A set of data (predicates) on the input flows of \emph{System} is constrained via {\scriptsize $\ll$}Proof Assumption{\scriptsize $\gg$} blocks during proof construction. A set of proof obligations is constructed via a function $F$ block and the output of $F$ is specified as input to block $P$.
$P$ is connected to an {\scriptsize $\ll$}Assertion{\scriptsize $\gg$} block and returns \emph{true} when the predicates set on the input data flows of the outline model is satisfied. Whenever {\scriptsize $\ll$}Assertion{\scriptsize $\gg$} is utilized, \sdv\ verifies if the specified input data flow is always \emph{true}.
The underlying \prover\ engine allows the formal verification of properties for a given model. Any failed proof attempt ends in the generation of a counterexample representing an execution path to an invalid state. A harness model is generated to analyze the counterexample and refine the model.

\section{\smc}
\smc\ performs the probabilistic analysis of properties by monitoring simulations of the complex hybrid system in a given stochastic environment and using results from the statistics to decide if the system satisfies the property with some degree of confidence. It represents systems via networks of Timed Automata (TA) \cite{alurTA} whose behaviors depend on both stochastic and non-linear dynamical features.
Its clocks evolve with various rates, which are specified with ordinary differential equations. \smc\ provides a number of queries related to the stochastic interpretation of TA (STA) \cite{david2015uppaal} and they are as follows, where $N$ and $bound$ indicate the number of simulations to be performed  and the time bound on the simulations respectively:

\begin{itemize}
\item \emph{Probability Estimation} estimates the probability of a requirement property $\phi$ being satisfied for a given STA model within the time bound: $Pr[bound]$ $\phi$;
\item \emph{Hypothesis Testing} checks if the probability of $\phi$ satisfied within a certain probability $P_0$: $Pr[bound]$ $\phi$ $\ge$ $P_0$;
\item \emph{Simulations}: \smc\ runs multiple simulations on the STA model and the $k$ (state-based) properties/expressions $\phi_1,..., \phi_k$ are monitored and visualized along the simulations: $simulate$ $N$ $[\leq$ $bound]\{\phi_1,..., \phi_k\}$;
\item \emph{Probability Comparison} compares probabilities of two properties being satisfied in certain time bounds: $Pr[bound_1]$ $\phi_1$ $\ge$ $Pr[bound_2]$ $\phi_2$
\item \emph{Expected Value} evaluates the maximal or minimal value of a clock or an integer value while \smc\ checks the STA model: $E[bound; N](min:\phi)$ or $E[bound; N](max:\phi)$.
\end{itemize}

\chapter{Running Example}
\label{sec:case-study}
\begin{figure}[htbp]
  \centering
  \includegraphics[width=4in]{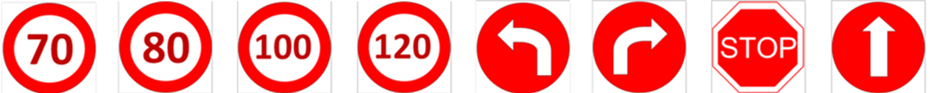}
  \caption{The eight types of traffic signs recognized by \av}
  \label{fig_traffic_sign}
\end{figure}

An autonomous vehicle application using Traffic Sign Recognition (\av) is adopted to illustrate our approach. The \av\ reads the road signs, e.g., ``speed limit'' or ``right/left turn ahead'', adjusts speeds and movements accordingly. The \av\ example is a 51Arduino-DS Robot vehicle consisting of a Robot-Eyes camera with  640*480p resolution, an 1800mah lithium battery, and four wheels and four motors.  \av\ is able to recognize eight sign types (see Fig.\ref{fig_traffic_sign}) automatically and restrict the speed of four wheels based on the signs. We manually set a max/min speed range (\{100,120\}/\{70,80\}). The structure of \av\ functionality is viewed as the Functional Design Architecture (FDA) in \ed\  augmented with timing/energy constraint: The { \gt{Camera}} captures sign images and sends the image to { \gt{Sign Recognition}} periodically.  { \gt{Sign Recognition}} analyzes each frame of the detected images in the YCbCr color space \cite{YCbCr} and compute the desired images. { \gt{Controller}} decides how the speed of the vehicle is adjusted based on the images and the current speeds of vehicle then sends the speed change requests to { \gt{Speed Calculator}} accordingly in which the speeds of wheels are computed and changed.

 \

\begin{figure}[htbp]
  \centering
  \includegraphics[width=6.5in]{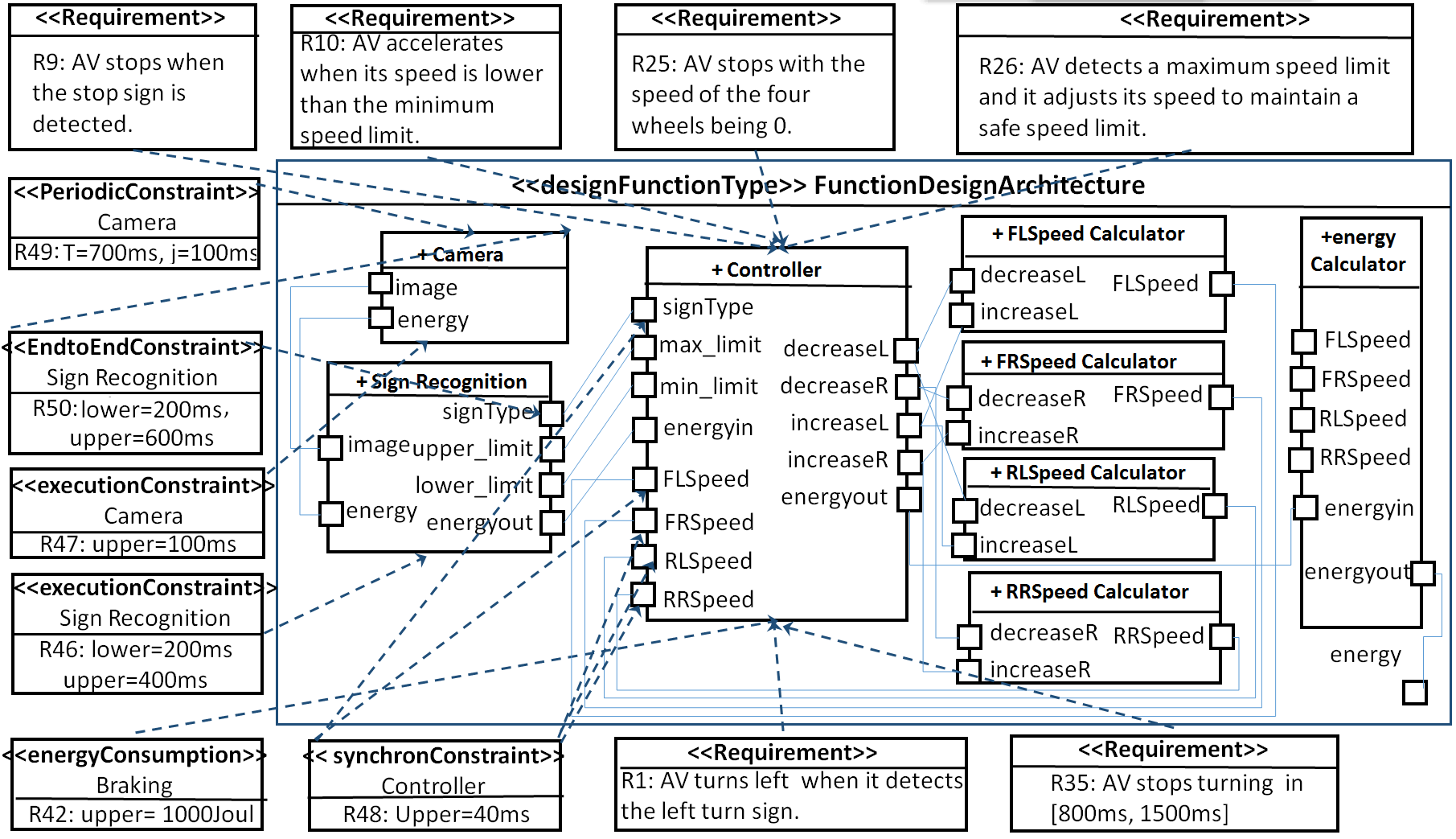}
  \caption{Design level \ed~model of \av}
  \label{fig:East-adl model}
\end{figure}

The requirements are listed as follows:
R1 to R36 are given as functional properties and R37 to R45 are requirements related to energy consumption, i.e., { \gt{Energy}} constraints. We also consider { \gt{Delay, Synchronization, Repeat}} timing requirements (R46 to R51) illustrated on top of the \av\ \ed\ model, which are sufficient to capture the constraints described in Fig.\ref{fig:East-adl model}. { \gt{Energy}} constraint refers to battery power consumption of \av.

\begin{itemize}[]
\item R1. When the vehicle detects a left turn sign and it is going straight with a constant speed, it must turn towards left;
\item R2. When the vehicle detects a left turn sign and it is  accelerating, it must turn towards left;
\item R3. When the vehicle recognizes a left turn sign and it is decelerating, it must turn towards left;
\item R4. When the vehicle detects a right turn sign and it is going straight with a constant velocity, it must turn towards right;
\item R5. When the vehicle recognizes a right turn sign and it is accelerating, it must turn towards right;
\item R6. When the vehicle recognizes a right turn sign and it is decelerating, it must turn towards right;
\item R7. When the vehicle detects a stop sign and it is going straight with a constant velocity, it must start to brake;
\item R8. When the vehicle detects a stop sign and it is accelerating, it must start to brake;
\item R9. When the vehicle detects a stop sign and it is decelerating, it must start to brake;
\item R10. If the vehicle recognizes a minimum speed limit sign (70 or 80) when going straight with a constant speed, and the current velocity of the vehicle is smaller than the speed limit, the vehicle should accelerate;
\item R11. If the vehicle recognizes a maximum speed limit sign (100 or 120) when going straight with a constant speed, and the current velocity of the vehicle is greater than the speed limit, the vehicle should decelerate;
\item R12. If the vehicle recognizes a minimum speed limit sign (70 or 80) when it is decelerating, and the current speed of the vehicle is smaller than the speed limit, the vehicle should accelerate;
\item R13. If the vehicle detects a minimum speed limit sign (70 or 80) when going straight with a constant speed and the current velocity of the vehicle is greater than the speed limit, it should maintain its speed;
\item R14. If the vehicle detects a maximum speed limit sign (100 or 120) when accelerating, and the current velocity of the vehicle is greater than the limit, it must decelerate;
\item R15. If the vehicle detects a maximum speed limit sign (100 or 120) when going straight with a constant velocity and its current velocity is smaller than the speed limit, the vehicle will maintain its speed;
\item R16. When the vehicle detects a stop sign and it is turning left, it should decrease its speed and stop (the speeds of the four wheels should decelerate to 0);
\item R17. When the vehicle detects a stop sign and it is turning right, it should decrease its speed and stop (the speeds of the four wheels should decelerate to 0);
\item R18. When the vehicle recognizes a go straight sign and it is accelerating its speed, it should continue to accelerate.
\item R19. When the vehicle recognizes a go straight sign and it is decelerating its speed, it should continue to decelerate;
\item R20. When the vehicle is running at a high speed (i.e., the speed is greater than or equal to 70 m/s) and it detects a left turn sign, it should decelerate the speeds of the left wheels (in the front and rear) to turn towards left;
\item R21. When the vehicle is running at a low speed (i.e., the speed of the vehicle is less than 70 m/s) and it detects a left turn sign, it will accelerate the speeds of the right wheels to turn towards left;
\item R22. When the vehicle is running at a high speed  and it detects a right turn sign, it should decelerates speeds of the  right wheels to turn right;
\item R23. When the vehicle is running at a low speed and it detects a right turn sign, it will accelerate the speeds of the  left wheels to turn right;
\item R24. When the vehicle is braking, it should finally stop with speeds of the four wheels 0;
\item R25. When the vehicle decelerate to a full stop, the speeds of the four wheels should be 0;
\item R26.When the vehicle detects a maximum speed limit sign with 100 as the limit, it automatically adjusts its speed to maintain a \emph{safe limit speed}.
\item R27.When the vehicle detects a maximum speed limit sign with 120 as the limit, it automatically adjusts its speed to maintain a \emph{safe limit speed}.
\item R28.When the vehicle detects a minimum speed limit sign with 70 as the limit, it automatically adjusts its speed to maintain a \emph{safe limit speed}.
\item R29.When the vehicle detects a maximum speed limit sign with 80 as the limit, it automatically adjusts its speed to maintain a \emph{safe limit speed}.
\item R30. When the vehicle is turning towards left, the speed of the wheels on the left side should be not greater than the speed of the wheels on the right side;
\item R31. When the vehicle is turning towards right, the speed of the wheels on the right should be not greater than the speed of the wheels on the left side;
\item R32. Once the vehicle starts to accelerate, it must complete its acceleration within  certain time interval [0ms, 2400ms];
\item R33. Once the vehicle starts to decelerate, it must complete its deceleration within certain time interval [0ms, 2400ms];
\item R34. When the vehicle detects a stop sign, it must stop (the speeds of the four wheels become 0) within a certain time interval [0ms, 1200ms];
\item R35. If the vehicle detects a left turn sign, it must turn left within a certain time interval [800ms, 1500ms];
\item R36. If the vehicle detects a right turn sign, it must turn right within a certain time interval [800ms, 1500ms];
\item R37. {  \gt {Energy constraint}}: The battery {  \gt{energy}} consumption for camera to captures an image should be within [1, 3] Joul;
\item R38. {  \gt {Energy constraint}}: The battery {  \gt{energy}} consumption for computing the desired image (sign type) of an captured image in {  \gt{Sign Recognition}} should be within a certain interval [1, 5] Joul;
\item R39. {  \gt {Energy constraint}}: The battery {  \gt{energy}} consumption for the vehicle to maintain a constant speed to go straight should be equal to the speed of the vehicle;
\item R40. {  \gt {Energy constraint}}: The battery {  \gt{energy}} consumption for the vehicle to turn left should be within a certain interval [1, 270] Joul.
\item R41. {  \gt {Energy constraint}}: The battery {  \gt{energy}} consumption for the vehicle to turn right should be within a certain interval [1, 270] Joul.
\item R42. {  \gt {Energy constraint}}: The battery {  \gt{energy}} consumption for the vehicle to stop should be within a certain interval [1, 1000] Joul.
\item R43. {  \gt {Energy constraint}}: The battery {  \gt{energy}} consumption for the vehicle to accelerate should be within a certain interval [0, 400] Joul.
\item R44. {  \gt {Energy constraint}}: The battery consumption for the vehicle to decelerate should be within a certain interval [0, 400] Joul.
\item R45. {  \gt {Energy constraint}}: When the velocity of vehicle is 0, the vehicle consumes no kinetics energy.
\item R46. A {  \gt{Delay}} constraint applied on {  \gt{Sign Recognition}} \fp\ is between 200 ms and 400 ms;
\item R47. A {  \gt{Delay}} constraint applied on {  \gt{Camera}} is within 100 ms;
\item R48. {  \gt{Synchronization}} constraint: The recognized sign types and the speeds of front/rear left/right wheels ({  \gt{signType}} and {  \gt{F(R)L(R)Speed}} ports on {  \gt{Controller}}) must be detected by {  \gt{Controller}} within a given time window, i.e., the tolerated maximum constraint = 40 ms;
\item R49. A {  \gt{Periodic}} acquisition of {  \gt{Camera}} must be vehicle should be carried out for every 700 ms with a jitter 100 ms.
\item R50. A {  \gt{Delay}} is measured from {  \gt{Camera}} to {  \gt{Sign Type}}. The time interval is bounded with a minimum value of 200 ms and a maximum value of 600 ms;
\item R51. The {  \gt{Delay}} from {  \gt{Camera}} to {  \gt{Sign Type}} should be less than or equal to the sum of the worst execution times of {  \gt{Camera}} and {  \gt{Sign Recognition}}.

\end{itemize}
According to the \ed\ meta-model, the timing constraint describes a
design constraint, but has the role of a property, requirement or
validation result, based on its {  \gt{Context}} \cite{EAST-ADL}.
The \tdl2\ meta-model is integrated with the \ed\ meta-model and is
supplemented with structural concepts from \ed. The \ed/\tdl2~constraints
contain the identifiable state changes as \emph{Events}. The
causality related events are contained as a pair by
\emph{EventChains}. Based on \emph{Event} and \emph{EventChains}, data dependencies, control flows, and critical execution paths are
represented as additional constraints for the \ed\ functional
architectural model, and apply timing constraints on these paths.

{  \gt{Delay}} constraint gives duration bounds (minimum and maximum)
between two events \emph{source} and \emph{target}, i.e., period,
end-to-end delays.  This is specified using \emph{lower, upper}
values given as either {  \gt{Execution}} constraint (R46, R47) or {  \gt{End-to-End}} constraint (R50). {  \gt{Synchronization}} constraint describes how tightly the
occurrences of a group of events follow each other. All events must
occur within a sliding window, specified by the \emph{tolerance}
attribute, i.e., a maximum time interval allowed between events (R48). { \gt{Repeat}} constraint states that the period of the successive
occurrences of a single event must have a length of at least a
\emph{lower} and at most an \emph{upper} time interval. The interval
is given as { \gt{Periodic}} constraint (R49). Those non-functional requirements (properties) are formally specified and verified in our approach using various analysis techniques that are described further in the following chapters.

\chapter{Modeling and Translation of \ed~Nonfunctional Properties in \simu~\& \staf}
\label{sec:ss-trans}
\section{Architectural Mapping to \simu~\& \staf~ using \me}
\me~ \cite{metaedit} tool provides \emph{.MDL\ export} option to automatically generate \emph{MDL} files from \ed~ model.  The generated file of the \ed~model is an MDL file consisting of eight {\tiny$<<$ }{Model Reference}{\tiny $>>$} blocks inside. The \emph{MDL} file is mapped to the  { \gt{FunctionDesignArchitecture}} \ft~while each {\tiny$<<$ }{Model Reference}{\tiny$>>$ } blocks correspond to the \fp\ with the identical name respectively.

\begin{figure}[htbp]
  \centering
  \includegraphics[width=6.5in]{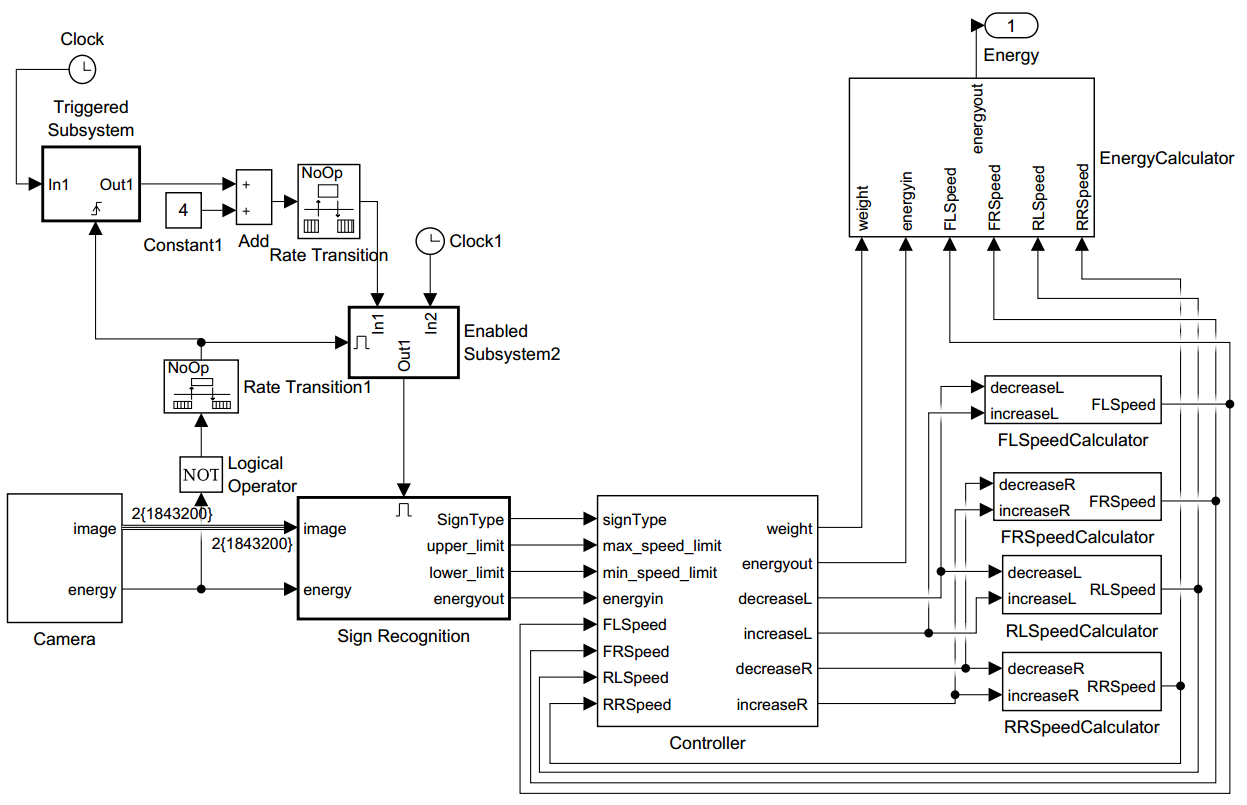}
  \caption{Simulink \& Stateflow model of \av}
  \label{fig:ss}
\end{figure}

\begin{figure}[htbp]
  \centering
  \includegraphics[width=3.5in]{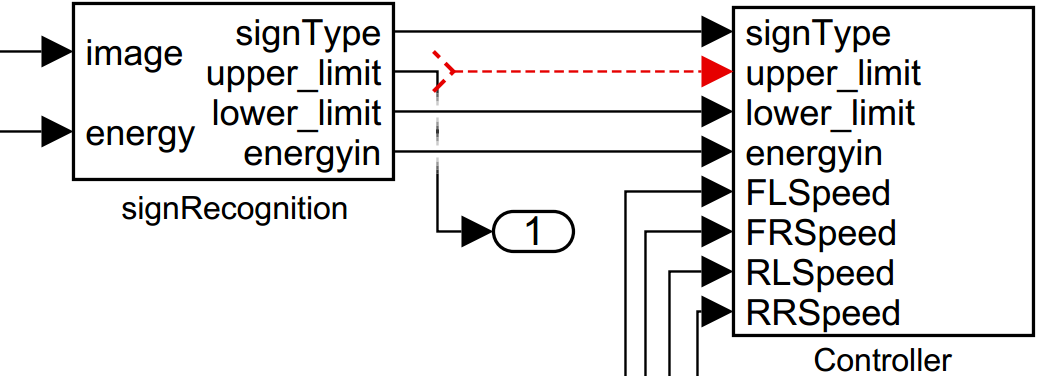}
  \caption{Error in generated model}
  \label{fig:error}
\end{figure}

The automatic model transformation is of great convenience for mapping architectural \ed~to Simulink/Stateflow model. However, if there are more than one connections from one output port in  \ed, after transformation, one of the connections may be lost in the exported model, which requests manual correction. As Fig.\ref{fig:error} shows, the connection from \emph{upper\_limit} port to { \gt{Controller}} is lost since there are two connections from the output port \emph{upper\_limit}.

\section{Behavioral Modeling of \fp s in \simu~\& \staf~}
We refine the model and construct the behaviour of \fp s with Stateflow chart and Simulink blocks. { \gt{Controller}} \fp~is modeled as a Stateflow chart while { \gt{Camera}}, { \gt{SignRecognition}}, { \gt{Energy Calculator}} and { \gt{(FL, FR, RL, RR) SpeedCalculator}} are modeled as subsystems with a number of descriptive blocks provided in Simulink block library. The architecture of Simulink/Stateflow model can be seen in Fig.\ref{fig:ss}.

\begin{figure}[htbp]
  \centering
  \includegraphics[width=6.5in]{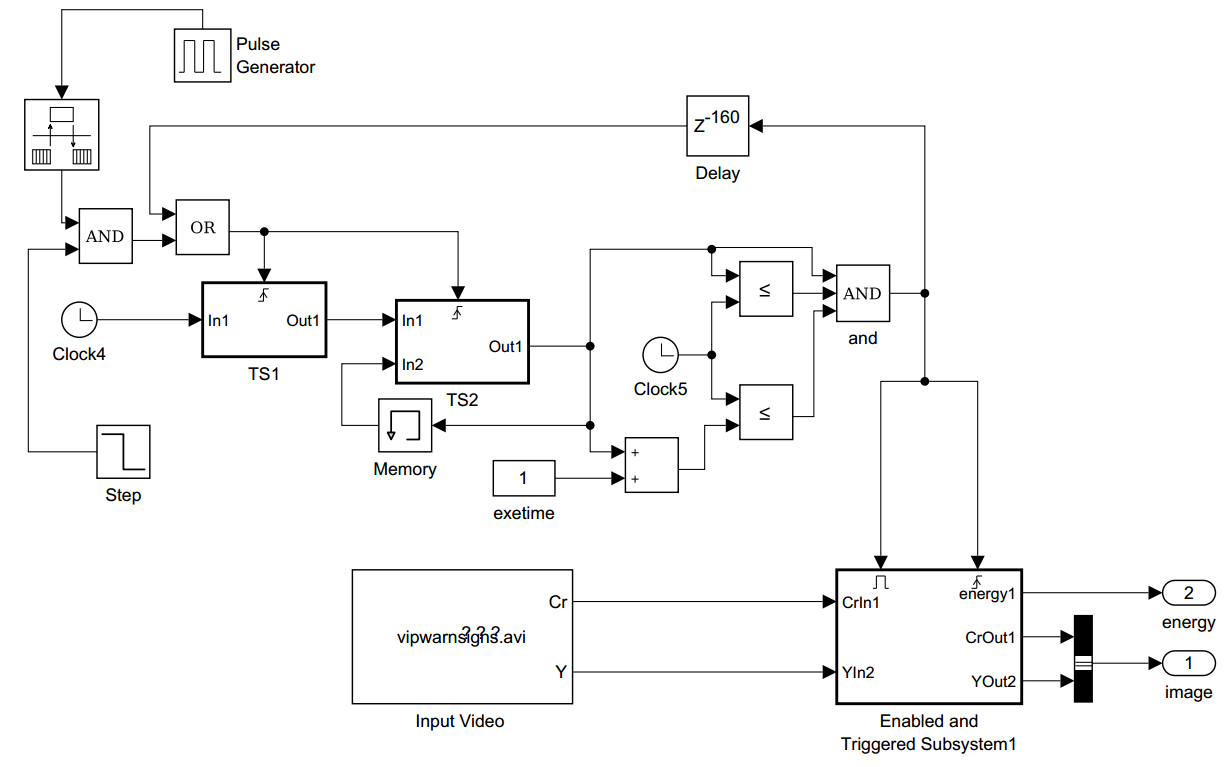}
  \caption{Simulink subsystem of Camera \fp}
  \label{fig:camera}
\end{figure}

As shown in Fig.\ref{fig:camera}, in { \gt{Camera}} subsystem, a video is loaded from the workspace by using a {\tiny$<<$ }{Multimedia File}{\tiny$>>$ } block. Then { \gt{Camera}} divides the source video into video frames (i.e. images) and output the frames to { \gt{SignRecognition}}.

\begin{figure}[htbp]
  \centering
  \includegraphics[width=6.5in]{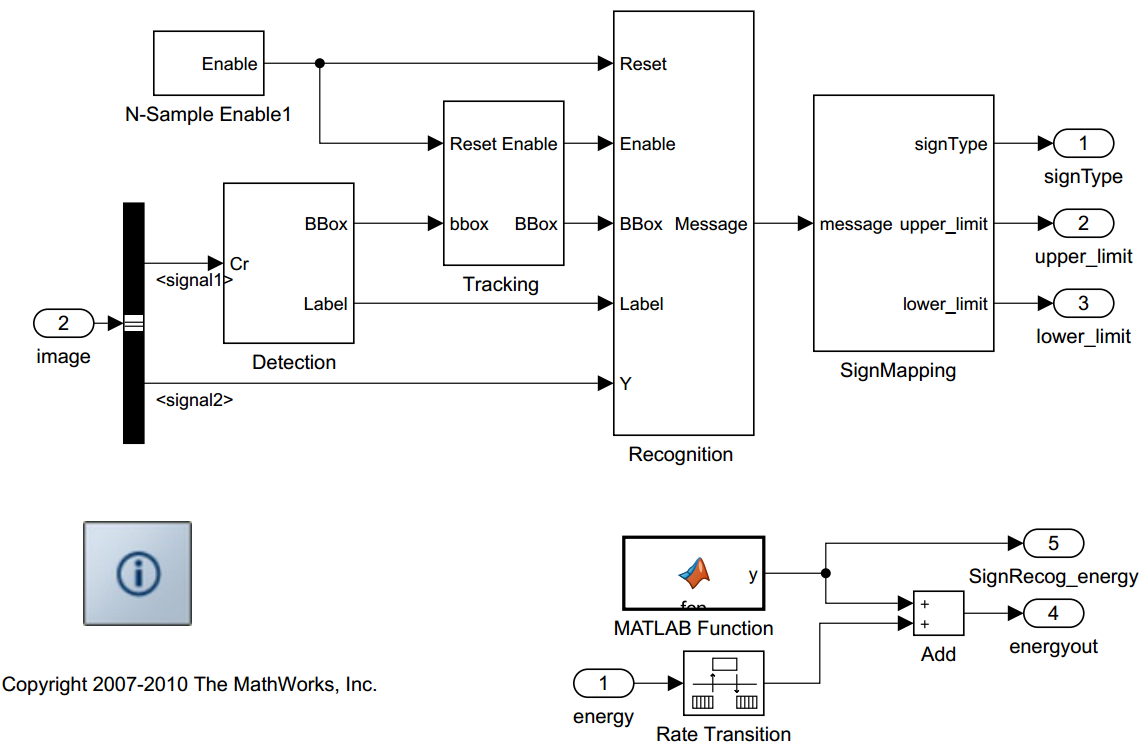}
  \caption{Simulink subsystem of SignRecognition \fp}
  \label{fig:signrecog}
\end{figure}

Fig. \ref{fig:signrecog} illustrates the structure of { \gt{SignRecognition}} subsystem, where the existed model for traffic warning sign recognition \cite{trafficwarningsign} provided by the Mathworks is applied and adapted. The model analyzes each input video frame in YCbCr color space. Y is the luma component and Cb and Cr are the blue-difference and red-difference saturability components. By thresholding and performing image processing on the Cr channel, the example extracts the outline of the key feature of the sign on the video frame. The model then compares the outline with each template signs stored in the \mt~workspace. If the outline is similar to any of the template signs, the model considers the most similar template to be the actual traffic sign.

\begin{figure}[htbp]
  \centering
  \includegraphics[width=6.5in]{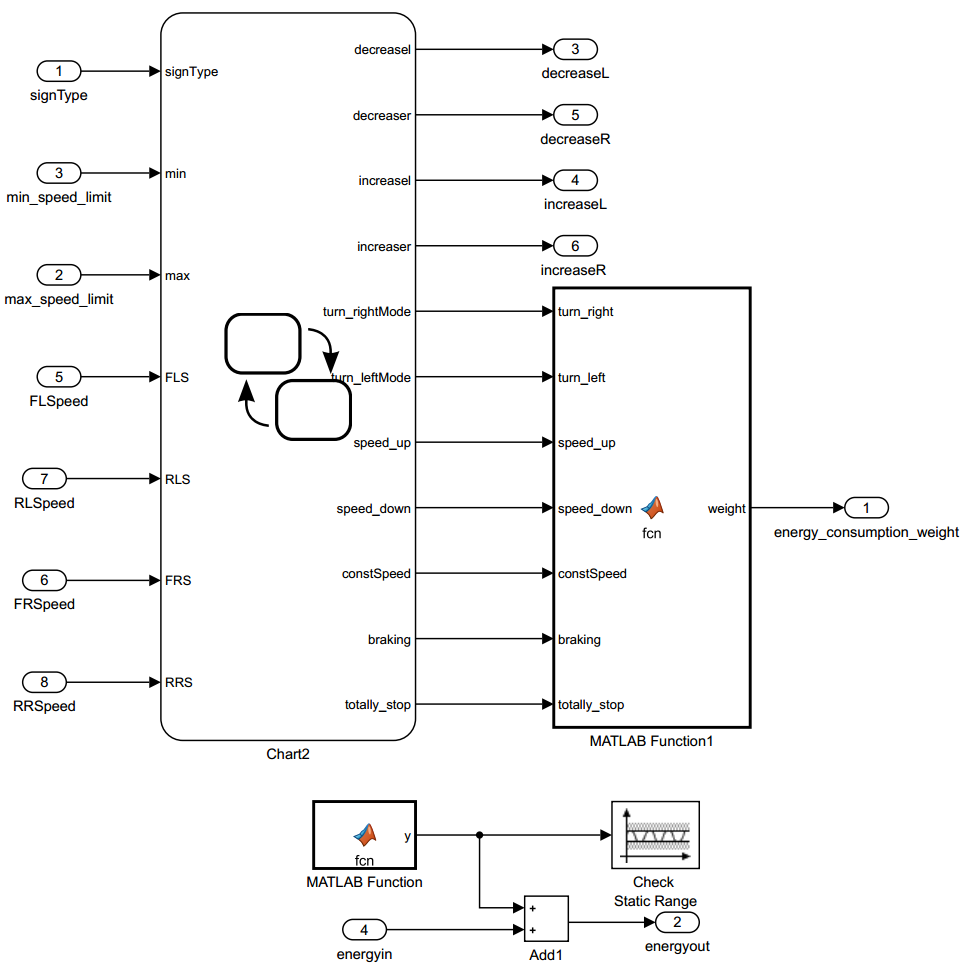}
  \caption{Simulink subsystem of Controller \fp}
  \label{fig:controlsub}
\end{figure}

\begin{figure*}[htbp]
\centering
\includegraphics[width=6.5in]{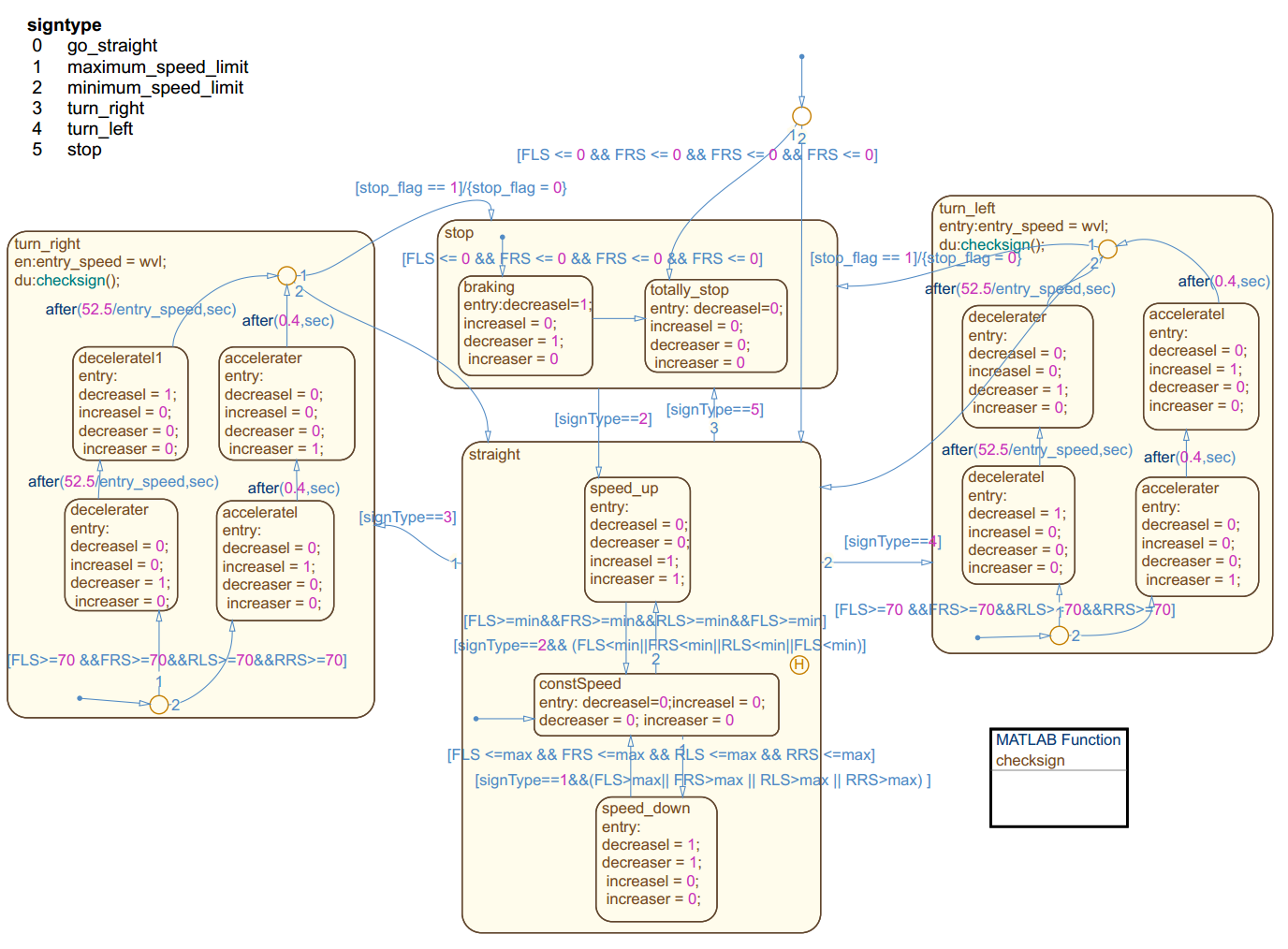}
\caption{The Stateflow of { \gt{Controller}} \fp}
\label{figure_controller}
\end{figure*}

{ \gt{Controller}} subsystem is presented in Fig. \ref{fig:controlsub} and the innner behaviour of { \gt{Controller}} is illustrated in the Stateflow chart in Fig.\ref{figure_controller}. The chart consists of four superstates \emph{turn\_right}, \emph{turn\_left}, \emph{straight}, \emph{stop} together with their child states. The edges represent transitions between the states with the conditions on the edges. The chart transits to either straight or stop based on the initial speeds of the wheels. The transitions will be taken according to the current speed of the wheels and the value of \emph{SignType}. For example, when the vehicle is in \emph{speed\_up} state and the detected \emph{SignType} is 5 (stop), \emph{braking} and \emph{stop} will be activated and \emph{decreaseL} and \emph{decreaseR} will become 1, indicating that the vehicle should decelerate the speeds of left wheels and right wheels. Whenever the vehicle detects a stop sign, the vehicle will start to brake. If the straight sign is detected, the vehicle will maintain the state/speed. If the vehicle recognizes a left turn sign, it can either decrease the speeds of the left wheels (in the  front and rear) or increase the speed of right wheels to turn, which depends on whether the speed of the vehicle is greater than 70m/s. After turning left (right), the vehicle will finally go straight and maintain the speed.
The embedded Matlab function \emph{checksign} is applied to detect whether the stop sign is recognized when the vehicle is in the turning (left or right) mode.

\begin{figure}[htbp]
  \centering
  \includegraphics[width=6in]{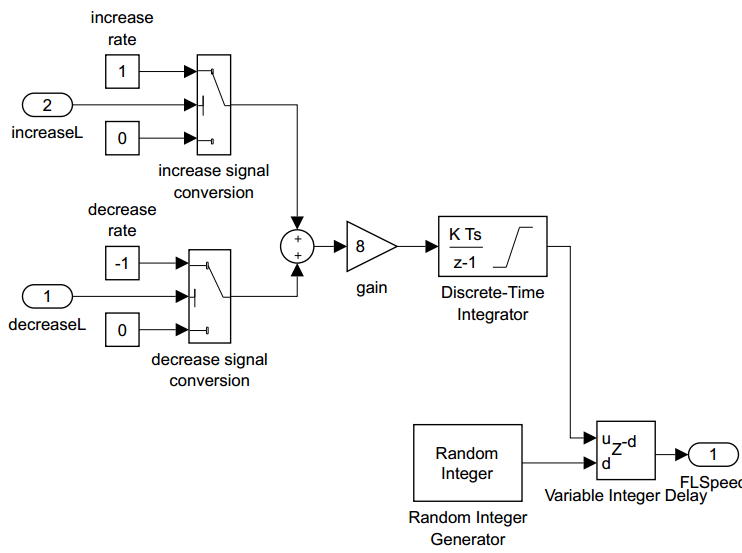}
  \caption{Simulink subsystem of FLSpeedCalculator \fp}
  \label{fig:flscal}
\end{figure}

{ \gt{SpeedCalculator}} subsystem calculates the speed of left (right) wheels. As shown in Fig. \ref{fig:flscal}, in the { \gt{FLSpeedCalculator}} subsystem, {\tiny$<<$ }{Discrete-time\ Integrator}{\tiny$>>$ } block and {\tiny$<<$ }{Gain}{\tiny$>>$ } operator are used to model the proportional relation between acceleration and speed. The value of the gain represents the acceleration of the movement (here is 8m/s$^2$). The speed (of the front/rear left/right wheels ) is calculated based on the values of four boolean variables: \emph{increaseL}, \emph{decreaseL}, \emph{increaseR} and \emph{decreaseR}, whose value determine whether the acceleration is positive or negative. The same pattern can be applied for { \gt{FRSpeedCalculator}}, { \gt{RLSpeedCalculator}} and { \gt{RRSpeedCalculator}} subsystems.

\section{Timing and Energy Constraints Translation}
\begin{figure}[htbp]
  \centering
  \includegraphics[width=2.4in]{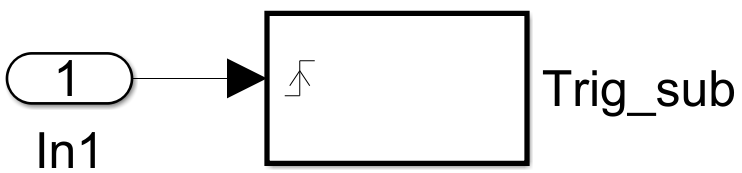}
  \caption{{ \gt{Event triggered}} timing constraint translation pattern}
  \label{fig_event_trigger_pattern}
\end{figure}

Timing and energy constraints in \ed\ are modeled by means of constraints specified on events and event chains. We show how the \ed\ constraints can be interpreted in \simu/\staf~(S/S) and provide corresponding modeling extensions in \s. We focus on { \gt{Synchronization, Execution, End-to-End, Periodic}} constraints (R46 to 51) and energy consumption constraints (R42) that are associated to \fp s in \ed. Additionally, we provide sufficient means to specify \emph{trigger} conditions for:

\begin{enumerate}
\item { \gt{Time-triggered}} \fp\ is triggered every period \emph{T}. We use \emph{sample time} and set \emph{sample time = T} for the block which corresponds to the \fp\ in \s. For example, { \gt{Camera}} \fp\ takes a picture every 20ms, the \emph{sample time} of { \gt{Camera}} block is set to 20ms, i.e., the block is executed every 20ms.

\item { \gt{Event-triggered}} \fp. The condition is expressed by a set of ports, i.e., at least one data/event must be received/occured on each port after the last \fp\ execution in order to trigger the \fp\ again. It can be modeled as { \gt{Trig\_sub}} in \s\  (Fig.\ref{fig_event_trigger_pattern}). A trigger event is on an { \gt{event\_trig\_in}} port, i.e., whenever an event occurs, the input signal of { \gt{event\_trig\_in}} becomes true and { \gt{Trig\_sub}} is activated.

\end{enumerate}

\begin{figure}
\centering
\subfigure[{ \gt{Time-triggered}} $f_p$ Translation]{
\label{fig_exe}
\includegraphics[width=2.5in]{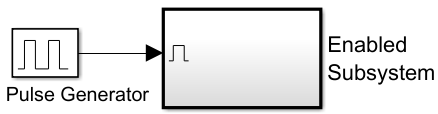}}
\subfigure[Signal from Pulse Generator]{
\label{fig_pulse}
\includegraphics[width=2in]{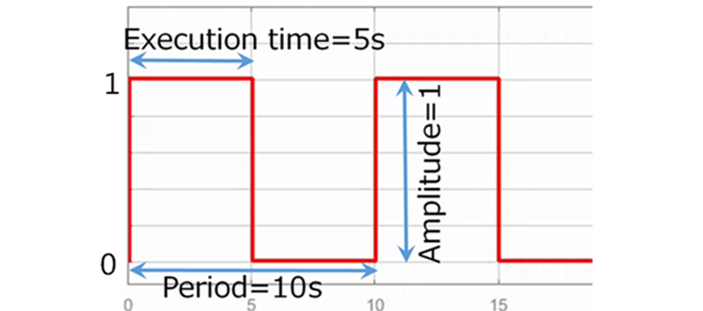}}
\subfigure[{ \gt{Event-triggered}} $f_p$ Translation]{
\label{fig:execpe}
\includegraphics[width=5.5in]{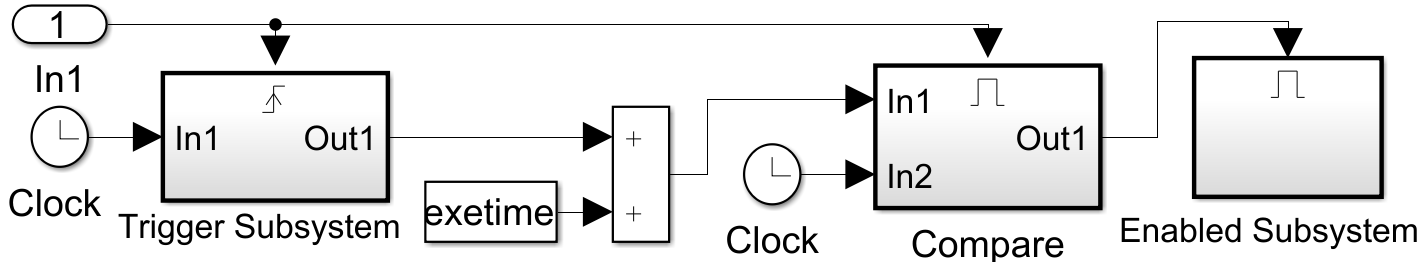}}
\caption{{ \gt{Execution}} timing constraint translation patterns}
\label{fig:exec-time-pattern}
\end{figure}

\noindent \textbf{Execution constraints}: In terms of { \gt{Time}}- or {\gt{Event-triggered}} \fp, two types of execution timing constraint translation patterns are provided: (1) { \gt{Time-triggered}} \fp\ translation pattern is shown in Fig.(a).  { \gt{Pulse Generator}} generates signals (square waves) based on \emph{Amplitude, Period, Pulse width} parameters (Fig.\ref{fig:exec-time-pattern}.(b)). \emph{Pulse width} is a duty cycle specified as a percentage of \emph{Period}. { \gt{Enabled Subsystem}} (representative of \fp) is executed when the triggering signal is greater than zero. Hence, the execution time of { \gt{Enabled Subsystem}} is $Pulse width * Period$ in this pattern; (2) { \gt{Event triggered}} \fp\ translation pattern is shown in Fig.\ref{fig:exec-time-pattern}.(c). { \gt{Trigg-}} {\gt{ered Subsystem}} is triggered by the rising edge of the square wave (changing curve from 0 to 1), where the rising edge indicates an event occurrence from { \gt{input}} port 1, { \gt{Clock}} outputs the current time instance, and $exetime$ is { \gt{Execution}} time constraint. For example, an event occurs and { \gt{Triggered Subsystem}} is triggered at the \emph{t} time point. In the meantime, the value of  { \gt{In1}} of { \gt{Compare}} becomes \emph{t+exetime}. If the current time instance is in [\emph{t}, \emph{t+exetime}] (i.e., the current time is within the range of execution timing constraint), { \gt{Out1}} becomes true and { \gt{Enabled Subsystem}} is executed.

\begin{figure}[htbp]
\centering
\subfigure[{\footnotesize \gt{Synchronization}} timing constraint translation pattern]{
\label{top}
\includegraphics[width=3.5in]{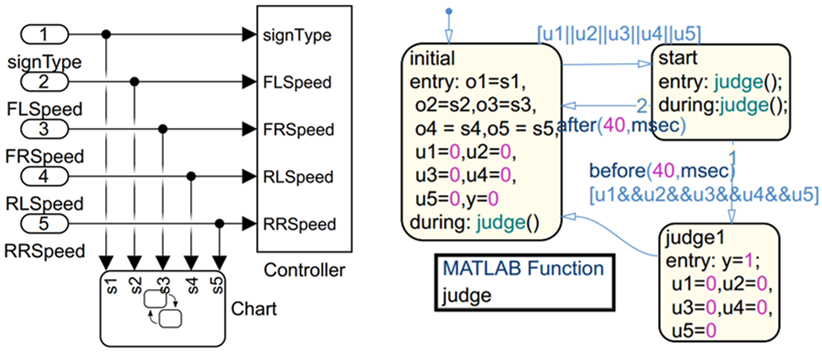}}
\subfigure[Signal examples]{
\label{example}
\includegraphics[width=2.5in]{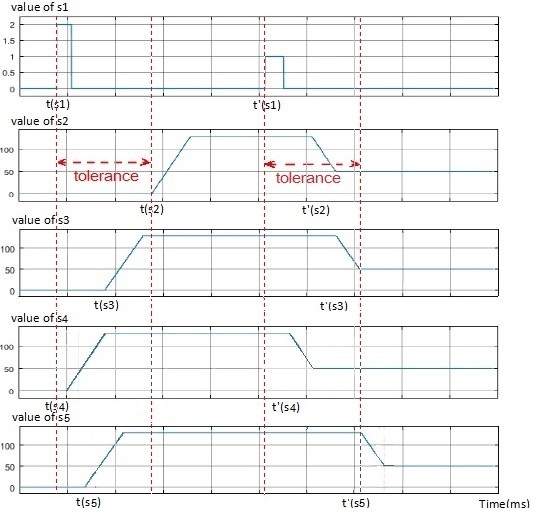}}
\caption{Synchronization timing constraint translation pattern and signal examples}
\label{fig:sync}
\end{figure}

\vspace{0.15in}
\noindent \textbf{Synchronization constraint} is a constraint on a set of events, which restricts the time duration among the $n^{th}$ occurrence of all the events in the set (i.e., maximum allowed time between the arrival of the event occurrences). The translation pattern for { \gt{Synchronization}} constraint is illustrated in Fig.\ref{fig:sync}. An observer \staf\ is used to record the exact arrival time of each event occurrence. A { \gt{Synchronization}} constraint (attached to the { \gt{Controller}} \fp\ in Fig.\ref{fig:East-adl model}) is the maximum tolerated time difference among the arrivals of recognized sign types and the speeds of left/right wheels (R48). To calculate the time interval  between the earliest and latest arrivals, the observer \staf\ ({\gt{Chart}}) is connected to { \gt{Controller}} in Fig.\ref{fig:sync}. The five inputs of { \gt{Chart}}, $s_1$ represents a parameter (signal) of a sign type, and $s_2, s_3, s_4, s_5$ denote the speed of the four wheels respectively. $o_i$ records the history value of each $s_i$, where $i = 1, 2, 3, 4, 5$. $u_i$ indicates if the signal has updated (changed).  A boolean \emph{y} indicates if the calculated time interval meets the { \gt{synchronization}} constraint. The arrival of each $s_i$ is monitored by comparing the current and previous values of the $s_i$  in { \gt{judge}} function. If any $s_i$ is updated, the { \gt{initial}} state is changed to { \gt{start}}. If the other remained signals are subsequently updated within the { \gt{synchronization}} constraint, \emph{y} is set to 1. A graphical illustration of $s_i$ is given in Fig.\ref{fig:sync}.(b), where, $t(s_i)$ is a time point of the $i^{th}$ occurrence of $s$. Similarly, a consecutive event of $s_i$ and its time point can be shown as $s'_i$ and $t'(s_i)$.  Recall R48, the difference among $t(s_i)$ and their consecutive events must be within the maximum tolerance (40ms).

\begin{figure}[htbp]
  \centering
  \includegraphics[width=5.5in]{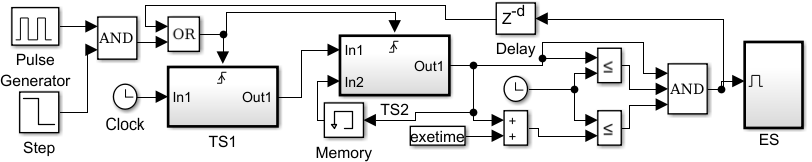}
  \caption{Periodic timing constraint translation pattern}
  \label{fig:pp}
\end{figure}

\vspace{0.05in}
\noindent \textbf{Periodic constraint} restricts the period of successive occurrences of \fp\ within a time interval [$T-j$, $T+j$], where $T$ and $j$ are a period and a jitter. The translation pattern is shown in Fig.\ref{fig:pp}: An enabled subsystem, { \gt{ES}}, which corresponds to \fp, is triggered by signals from two subsystems, { \gt{TS1}} and { \gt{TS2}}. $t_i$ denotes a time point of the $i^{th}$ triggered { \gt{ES}} (i.e., $i^{th}$ occurrence of \fp). To calculate  $t_1$, { \gt{Pulse Generator}} and { \gt{Step}} blocks trigger { \gt{TS1}} and { \gt{TS2}} at time \emph{T-j}. { \gt{TS1}} sends the current time \emph{T-j} to { \gt{TS2}}. Since \fp\ can be triggered at any time point in [-j, j], { \gt{TS2}} generates a value \emph{$r_i$} $\in$ [0, 2j]. According to the pattern, the $1^{st}$ occurrence of \fp\ happens at $t_1$ $=$ $T-j + r_1$, where $t_1$ $\in$ $[T-j, T+j]$. Similarly, the time point of each consecutive \fp\ occurrence ($t_2 = t_1 + T-j + r_2$, $t_3 = t_2+T+j+r_3$, etc.) can be calculated.  The  { \gt{Delay}} block  ensures that $t_i$ must be calculated prior to its lower bound ($t_{i-1}+T-j$).  In order to guarantee \fp\ is executed periodically, the execution timing constraint translation pattern is applied ({\gt{exetime}}). Using this pattern, R49 is validated in section \ref{sec:v-v}.

\vspace{0.05in}
\noindent \textbf{End-to-End constraint} specifies how long after the occurrence of a \emph{stimulus}, a corresponding \emph{response} must occur. The input and output of a block ($B$) in \s\ are triggered by sample time ($st$). Assume that $B_1$/$B_n$ is the first (\emph{stimulus})/last (\emph{response}) block of a series of connected blocks. $B_i$ is the $i^{th}$ block and its sample time is ${st}_i$. In Simulink, a time interval between $B_1$ (where the inputs arrive) and $B_n$ (from where the outputs leave) is $s_{sum}$ $=$ $\sum_{i=1}^ns_i$. In the case that $B_i$ finishes its execution before $B_i$ is triggered again (i.e., the execution time of $B_i$ is shorter than its triggered time interval), { \gt{End-to-End}} constraint (the sum of all execution times of each $B_i$) can be less than or equal to $s_{sum}$. Therefore, { \gt{End-to-End}} constraint cannot be expressed in \s\ using sample times. The following sections elucidate how this constraint can be modeled alternatively and analyzed.

\vspace{0.15in}
\noindent \textbf{Energy constraint} is associated to each \fp\ to restrict minimum/maximum resource allocation on the execution platform. The energy consumption of either the \fp\ or the whole systems consisting of \fp s is modeled using differential equation blocks in \simu. According to the various modes of \av, the amount of battery power expended for the mechanical motion of the wheels is calculated: \[energy = \int_0^t a * v dx\] where, $t$, $a$, and $v$ denote running time, coefficient reflects an energy rate associated to the current mode of \av, and wheel speed respectively.  Details of energy-aware analysis is presented in chapter \ref{sec:v-v}.

\chapter{Modeling and Translation of \ed~ Nonfunctional Properties in \smc}
\label{sec:smc-trans}
\section{Architectural and Behavioral of \ed~ and \simu~\& \staf~ to \uppaal~STA}
\begin{figure}[htbp]
  \centering
  \subfigure[TA]{
  \label{fig:automataofsimu}
  \includegraphics[width=1.3in]{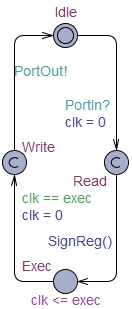}}
  \subfigure[Random generation of signType with probabilistic distribution]{
  \label{randomsign}
  \includegraphics[width=3.8in]{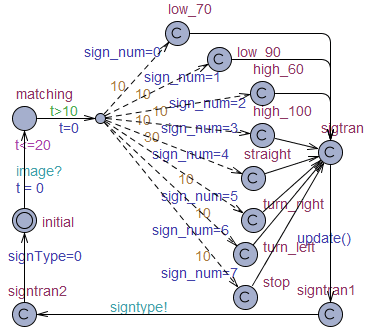}}
  \caption{Translate Simulink/Stateflow model to \uppaal\ STA}
  \label{template}
\end{figure}

\begin{figure}[htbp]
  \centering
  \subfigure[STA of {\gt{Camera}} \fp]{
  \label{camera_sta}
  \includegraphics[width=2.8in]{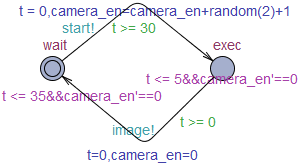}}
  \subfigure[STA of { \gt{SpeedCalculator}} \fp]{
  \label{speed_cal_sta}
  \includegraphics[width=3.6in]{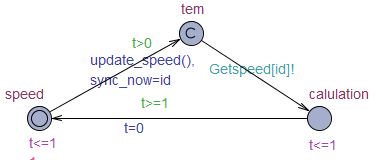}}
  \caption{Model the continuous behaviour of Simulink/Stateflow in \uppaal}
  \label{simulink_sta}
\end{figure}

To translate the dynamic and continuous behaviours of { \gt{Camera}} and { \gt{SignRcognition}} subsystems to \smc, we use and modify the interface automaton (Fig. \ref{fig:automataofsimu}) proposed in our previous work \cite{kang2016statistical}. \emph{Read} and \emph{Write} locations are committed to guarantee that there is no delay or interruption. Then, we capitalizes on the automatic code generation function in \mt~and generate c code using Simulink Coder. The code is modified and embedded in \emph{SignRec} function in the automaton. The automaton reads images data via \emph{PortIn?} channel.  After getting inputs, the automaton will execute and update the values of \emph{signType}, \emph{lower\_limit} and \emph{upper\_limit}. These parameters and signals are then written to output port, and the automaton will return to \emph{Idle} and wait for another trigger from the previous blocks. The STA shown in Fig.\ref{camera_sta} preserve the behaviour of { \gt{Camera}} \fp. Fig.\ref{speed_cal_sta} illustrates the STA that used for modeling the contimuous behaviour of  four { \gt{SpeedCalculator}} \fp s.

To model the stochastic behaviours of { \gt{SignRecognition}} \fp~that the traffic sign occurs randomly, a STA with eight non-deterministic edges is used (Fig.\ref{randomsign}). The eight edges corresponding to eight signs are associated with probability weight (here 0.3 for straight sign and 0.1 for the each of the other seven signs), which means that a discrete probabilistic choice ($\frac{30}{100}$ for straight sign versus $\frac{10}{100}$ for the other seven signs) is made.

\begin{figure}[htbp]
  \centering
  \includegraphics[width=6.5in]{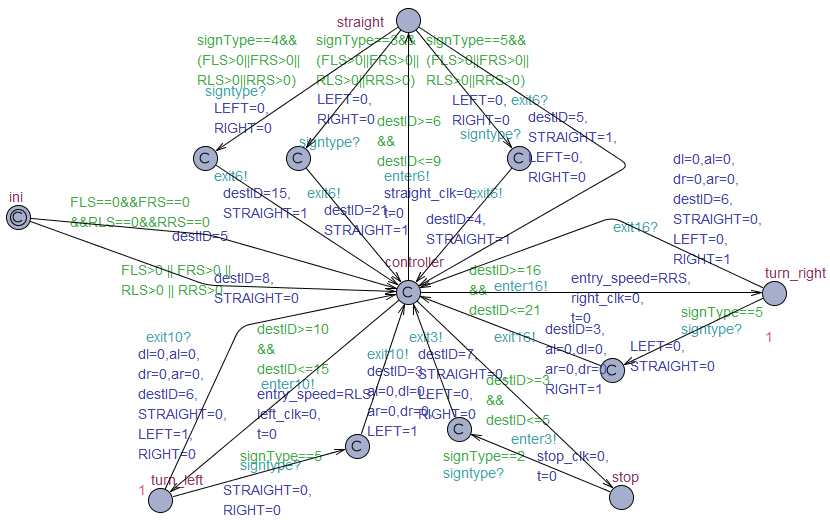}
  \caption{STA of \gt{Controller}}
  \label{fig:controller_sta}
\end{figure}

\begin{figure}[htbp]
  \centering
  \includegraphics[width=6.5in]{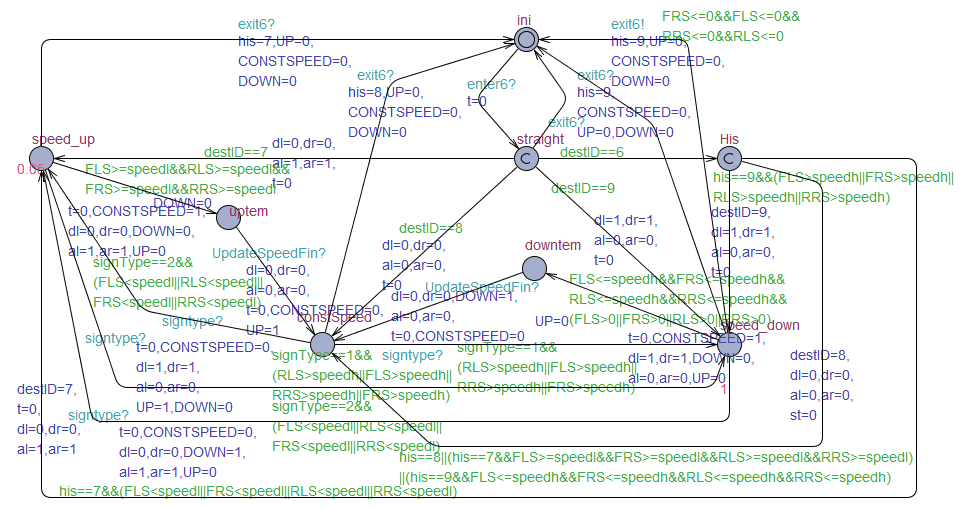}
  \caption{STA of \emph{straight} state in { \gt {Controller}}}
  \label{fig:straight_sta}
\end{figure}

\begin{figure}[htbp]
  \centering
  \subfigure[STA of \emph{turn\_left} state]{
  \label{turn_left_sta}
  \includegraphics[width=5.3in]{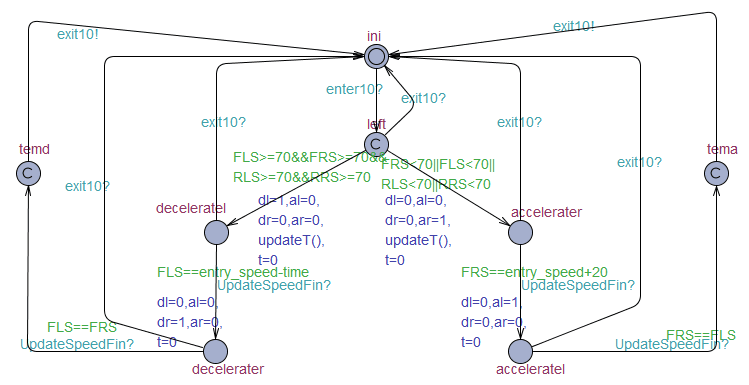}}
  \subfigure[STA of \emph{turn\_right} state]{
  \label{turn_right_sta}
  \includegraphics[width=5.3in]{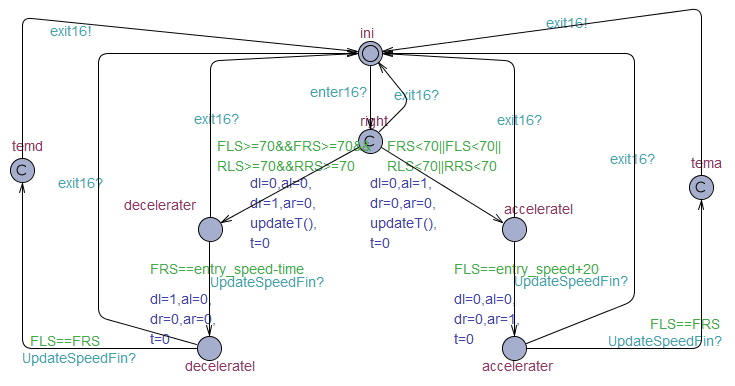}}
  \caption{STA of \emph{turn\_left} and \emph{turn\_right} state in {\gt {Controller}}}
  \label{fig_turn_template}
\end{figure}

\begin{figure}[htbp]
  \centering
  \includegraphics[width=2.6in]{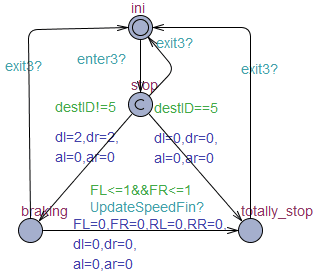}
  \caption{STA of \emph{stop} state in {\gt {Controller}}}
  \label{fig:stopsta}
\end{figure}

We translate the Stateflow chart into \smc\ STA by applying the methodology in \cite{kang2016statistical}. The \staf~chart is translated into 5 STAs: { \gt{Controller, Stop, Straight, Turn\_left}} and { \gt{Turn\_right}} respectively. { \gt{Controller}} STA corresponds to the topmost superstate of the Stateflow and it contains 4 locations: \emph{turn\_left/right}, \emph{stop}, \emph{straight}. Each location is mapped to the the STA with the identical name. As shown in Fig.\ref{fig:controller_sta} to Fig.\ref{fig:stopsta}, each STA has an \emph{initial} location. { \gt{Controller}} interacts with other STA (representatives of child states in Stateflow chart) via synchronization channels. For example, if {\gt{Controller}} stays in \emph{turn\_left} location, the \emph{initial} location of { \gt{Turn\_left}} STA  will be inactive. At the same time, the other three STAs, i.e. { \gt{Turn\_right}}, { \gt{Straight}} and { \gt{Stop}} STA  should stay in their \emph{initial} state respectively.


\section{Probabilistic Timing Constraints Translation }


We discuss semantics of the extended { \gt{Execution, Synchronization, Periodic}}, and { \gt{End-To-End}} timing constraints (\xtc) with probabilistic parameter according to \tdl2 \cite{TADL2}. Afterwards, we provide \xtc\ translation in STA  and discuss it with the view point of analysis engine \smc. \xtc\ and its translation follow the weakly-hard ({\gt{WH}}) approach \cite{TADL2}, which describes a bounded number of all event occurrences are allowed to violate the constraints within a time window. The semantics of the { \gt{WH(c,m,K)}} is that a system behavior must satisfy a given timing constraint { \gt{c}} at least { \gt{m}} times out of { \gt{k}} consecutive occurrences of events. We use object-oriented notation to define the attributes of occurrences of an event. These attributes are time points showing when instances of the event happens, e.g., { \gt{\fp.response}} refers to the time at which the response event of \fp\ is specified.



\begin{figure*}
\centering
\subfigure[{ \gt{Execution}}]{
\label{fig_execution}
\includegraphics[width=2in]{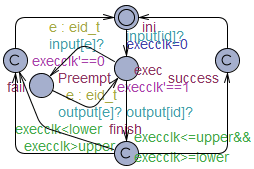}}
\subfigure[{ \gt{Synchronization}}]{
\label{fig_synch}
\includegraphics[width=3in]{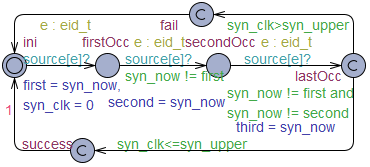}}
\subfigure[{ \gt{Periodic}}]{
\label{fig_periodic}
\includegraphics[width=1.5in]{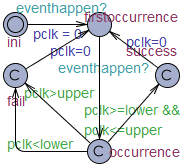}}
\subfigure[{ \gt{End-to-End}}]{
\label{fig_endtoend}
\includegraphics[width=1.5in]{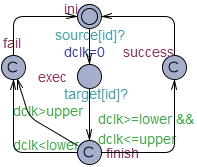}}
\caption{STA of \ed/\tdl2~Probabilistic Extension of Timing Constraints}
\label{fig:smctemplate}
\end{figure*}

\vspace{0.15in}
\noindent \textbf{{ \gt{WH(Execution,m,k)}}} limits the time between the starting and stopping of an executable \fp\ following \emph{run-to-complete} semantics, i.e., not counting the intervals when the executable \fp\ has been interrupted. At least { \gt{m}} times out of { \gt{k}} consecutive occurrences of the event, { \gt{\fp.start}}, satisfy: { \gt{lower}} $\leq$ $\mid${ \gt{\fp.stop}} $\setminus$ ({ \gt{\fp.preempt}}, { \gt{\fp.resume}})$\mid$ $\leq$ { \gt{upper}}, where { \gt{lower}} ({\gt{upper}}) denotes the time point showing when the \fp\ starts (stops). The STA modeling of \fp's execution constraint is shown in Fig.\ref{fig:smctemplate}.(a). The STA is executed via \emph{input[id]?} and its completion is updated via \emph{output[id]?}. The execution time of \fp\ (\emph{execclk}) is calculated in the \emph{exec} location. When \fp\ is started, a local clock \emph{execclk} is reset. If \fp\ is preempted (\emph{input[e]?}), \emph{execclk} holds (rate = 0) until the \fp\ is resumed. When \fp\ is completed,  STA moves to \emph{finish} location and it determines if \emph{execclk} is within the given { \gt{Execution}} constraint. In case of satisfying the constraint, it moves to \emph{success} location, otherwise stays in \emph{fail} location. The interpretation of { \gt{WH(Execution,m,k)}} is given as a \emph{Hypothesis Testing} query in \smc\ and it is as follows, $P$: $Pr[bound]$ ($[\ ]\neg$$STA_{Execution}.fail$) $\ge$ $P$, where \emph{bound} and $P$ indicate the time bound on the simulation and $\frac{m}{k}$ respectively.

\vspace{0.15in}
\noindent \textbf{{ \gt{WH(Synchronization,m,k)}}}: A set of input events of an \fp\ constrains the maximum allowed time delay attribute, \emph{tolerance}, among the arrival of the event occurrences. For any event { \gt{e}} in the set s.t. the fastest arrival event ({\gt{e(i).fastest}}) and the slowest arrival event ({\gt{e(j).slowest}}), where { \gt{i $\neq$ j}}, at least { \gt{m}} times out of { \gt{k}} consecutive occurrences of the set satisfy $\mid${ \gt{e(j).slowest-e(i).fastest}}$\mid$ $\leq$ {\gt{tolerance}}. The corresponding STA in Fig. \ref{fig:smctemplate}.(b) specifies the time width within which a set of ''input'' event should occur. For simplicity, we assume three input events occur (denoted \emph{source[e]?}). The parameter \emph{syn\_upper} (represented as \emph{tolerance}) determines the maximum timed allowed among the three inputs. { \gt{WH(Synchro-}} {\gt{nization,m,K)}} is specified as $P$: $Pr[bound]$ ($[\ ]$ $\neg$ $STA_{Synchronization}.fail$) $\ge$ $P$. Output synchronization constraint translation is similar to the input synchronization except that instead of ''input'', ''output'' are constrained to occurred in a specified time width.

\vspace{0.15in}
\noindent \textbf{{ \gt{WH(Periodic,m,k)}}} limits the period of the successive occurrences of a single event \fp\ including \emph{jitter}. For any { \gt{s(i)}} $\in$ { \gt{\fp.start}} s.t. its successive occurrence { \gt{s'(i)}} starting at { \gt{s(i)}}, out of { \gt{k}} consecutive occurrences { \gt{s(i)}}, i.e., { \gt{s(i)}} to { \gt{s(i+k)}}, at least { \gt{m}} sequences satisfy ({\emph{T-jitter}} $\leq$ { \gt{$\mid$s'(i)-s(i)$\mid$}} $\leq$ {\emph{T+jitter}}) $\land$ ({ \gt{lower}} $\leq$ { \emph{T}} $\leq$
{ \gt{upper}}), where { \emph{T}} denotes a bounded length of at least a \emph{lower} and at most an \emph{upper} time interval, given as the period without \emph{jitter}.  The STA in Fig. \ref{fig:smctemplate}.(c) enforces the event \emph{eventhappen} to arise within the  bounded time interval between every two consecutive occurrences of it.  When the $1^{st}$ occurrence happens, STA resets its local clock ($pclk$) and starts counting until the $2^{nd}$ occurrence happens. Afterwards, it judges if $pclk$ is within  the { \gt{periodic}} constraints and changes its current state to either (\emph{success} or \emph{fail}) based on the judgement. Finally, STA returns to \emph{firstocccurrence} and repeats the calculation on the next two consecutive occurrences of the event. { \gt{WH(End-to-End,m,k)}} is specified as $P$: $Pr[bound]$ ($[\ ]$ $\neg$$STA_{Periodic}.fail$) $\ge$ $P$.

\vspace{0.15in}
\noindent \textbf{{ \gt{WH(End-to-End,m,k)}}} limits { \gt{tolerance}} between the occurrences of two events (\fp s) \emph{source, target}. Only one-to-one occurrence patterns are allowed. For any {\gt{s}} $\in$ { \gt{\fp.source}} s.t. { \gt{s=source(i)}} for some { \gt{i}} $\leq$ { \gt{$\mid$source$\mid$}}, out of { \gt{k}} consecutive occurrences { \gt{s}} of { \gt{\fp.source}}, i.e., { \gt{source(i)}} to { \gt{source(i+k)}}, at least { \gt{m}} satisfy { \gt{lower}} $\leq$ { \gt{\fp.target(i)-s}} $\leq$ { \gt{upper}}, where { \gt{\fp.target(i)-s}} denotes the corresponding delay of { \gt{\fp.source(i)}}. The STA in Fig. \ref{fig:smctemplate}.(d) specifies { \gt{tolerance}} between { \gt{\fp.source}} and { \gt{\fp.target}}. \emph{dclk} counts the delay ({ \gt{tolerance}}) based on inputs via \emph{source[id]?} and \emph{target[id]?}. After allowing one-to-one source and target occurrence patterns, STA changes its state to \emph{finish} and decides its successor location either \emph{success} or \emph{fail} based on checking if \emph{dclk} is within the { \gt{End-to-End}} constraint. The interpretation of { \gt{WH(End-to-End,m,k)}} is given as a query, $P$: $Pr[bound]$ ($[\ ]$ $\neg$$STA_{End-to-End}.fail$) $\ge$ $P$.

\section{Energy-aware Modeling and Estimation}
\begin{figure}[ht!]
  \centering
  \includegraphics[width=6.5in]{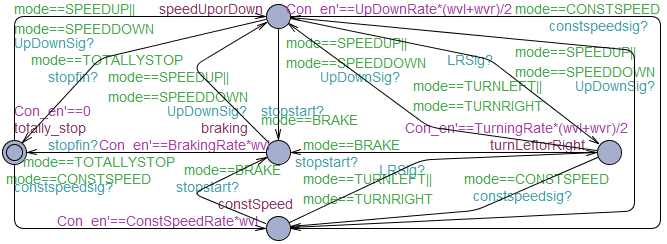}
  \caption{STA of the energy consumption of \gt{Controller}}
  \label{ENSTA}
\end{figure}

In order to estimate the energy consumption of hybrid \av\ based on \xtc\ and energy constraints, the { \gt{Controller}} \fp\ (which consumes batteries differently on variant modes) is selected and its ERT behaviors are formally specified in STA. The STA of { \gt{Controller}} (Fig.\ref{ENSTA}) presents stochastic hybrid behaviors with extended arithmetic on clocks and their rates: \emph{Con\_en} (defined as an \ode\ and assigned at each location). Different locations in the STA correspond to (one or two) states of { \gt{Controller}} (visualized in Stateflow in Fig.\ref{figure_controller}). A pair of modes having the same energy consumption rate can be combined into one and depicted as a location, i.e., since the energy consumption rates of { \gt{turn\_left/right}} modes are the same, they are expressed as a \emph{turnLeftorRight} location in STA. Because the energy consumption rate varies in different running mode, e.g., the energy is consumed fast when the vehicle is in braking mode and slowly in \emph{constSpeed} mode. We define \emph{Con\_en'} (the rate of \emph{Con\_en} ) with ordinary differential equation (ODE) and assign different values to \emph{Con\_en'} for different locations. \emph{UpDownRate}, \emph{BrakingRate}, \emph{ConstSpeedRate} and \emph{TurningRate} are user-supplied coefficients in ODE that represent battery consumption and its various rates on different modes respectively and \emph{BrakingRate $>$ UpDownRate $>$ TurningRate $>$ ConstSpeedRate}. The value of the coefficients indicate the rapidity of the energy consumption. The estimation of energy consumption requirement of Controller will be evaluated in chapter \ref{sec:v-v}.

\chapter{Experiments: Verification \& Validation}
\label{sec:energy}
\label{sec:v-v}

We employ \smc\ and SDV for the verification and \mt/\simu\ for the simulation of functional (R1--R36) and timing- and energy constraints properties (R37--R51).

\section{\sdv~verification}
\begin{figure}[htbp]
\centering
\subfigure[Input constraint]{
\label{fig_input}
\includegraphics[width=2in]{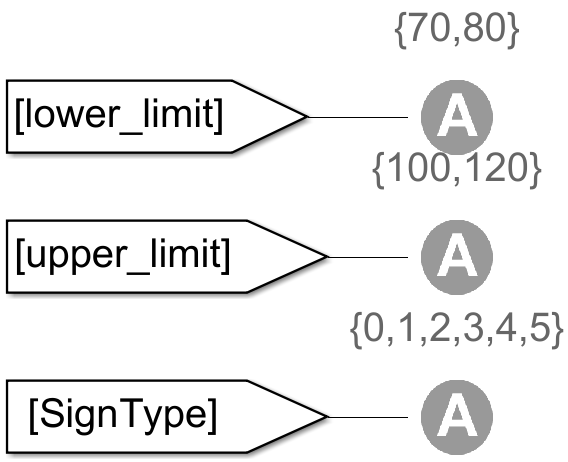}}
\subfigure[Model of requirement ]{
\label{fig_r1_model}
\includegraphics[width=4.5in]{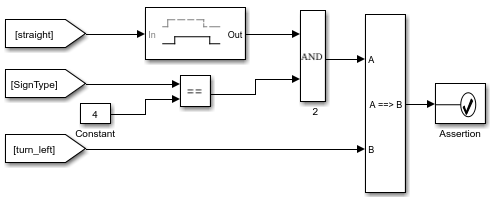}}
\caption{Example model of non-temporal functional property: If \emph{straight }state is active and \emph{SignType} is 4, \emph{turn\_left} state should be active next time step.}
\label{fig_func_model}
\end{figure}

Fig. \ref{fig_r1_model} illustrates an example model of non-temporal functional requirement in SDV. The value of \emph{SignType} signal ranges from 0 to 5, which represents straight sign, maximum/minimum speed limit sign, turn right/left sign and stop sign respectively. In Fig.\ref{fig_input}, \emph{lower\_limit}, \emph{upper\_limit} and \emph{SignType} are constrained in {\scriptsize$<<$ }{Proof Assumption}{\scriptsize$>>$ } block such that the signal must match one of the listed values at every time step, i.e., sample interval. The states in Stateflow chart are output as boolean signals for monitoring, e.g., the value of \emph{straight} is 1 when state \emph{straight} is active and 0 otherwise. A {\scriptsize$<<$ }{Detector}{\scriptsize$>>$ } block is applied to add a unit time delay for condition, suggesting that the condition should be true at previous time step.


\begin{figure}[htbp]
\centering
\subfigure[Model of requirement]{
\label{fig_r2_model}
\includegraphics[width=4in]{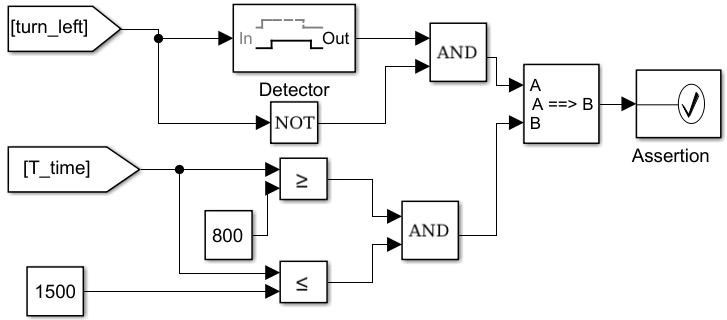}}
\subfigure[temporalCount]{
\label{fig_tempCount}
\includegraphics[width=2in]{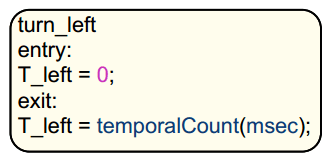}}
\caption{Example model of temporal functional property: If \emph{turn\_left} state is active at previous time step and inactive at current time step, \emph{T\_left} should be between 800 and 1500 (The time duration for turning left should be between 800 and 1500 ms).}
\label{fig_temp_model}
\end{figure}

The temporal functional requirement can be model as Fig. \ref{fig_temp_model}.
As shown in Fig.\ref{fig_tempCount}, once the chart exits \emph{turn\_left}, \emph{temporalCount(msec)} (which is an expression in absolute time temporal logic) returns the integer number of milliseconds that have elapsed since activation of the \emph{turn\_left} state (i.e, the time duration of turning left). Then the action assigns to \emph{T\_left}, which is an output variable to Simulink indicating the time duration of turning left.


The proof objective model of the requirement R7 is presented below.

\begin{figure}[htbp]
  \centering
  \includegraphics[width=4.5in]{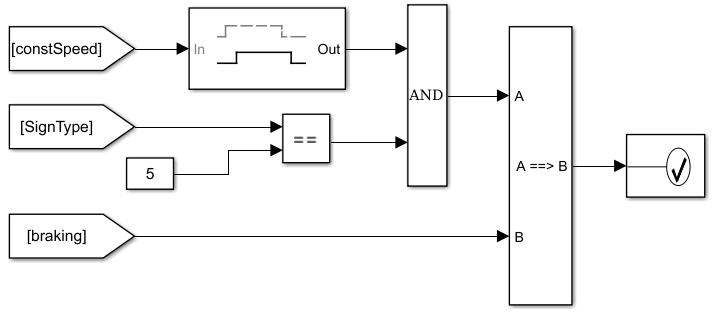}
  \caption{Model of requirement R7: If \emph{constSpeed} is active and the detected \emph{SignType} is 5, then the system will enter the \emph{braking} state in the next time step.}
  \label{R7}
\end{figure}

\begin{figure}[htbp]
  \centering
  \includegraphics[width=6.5in]{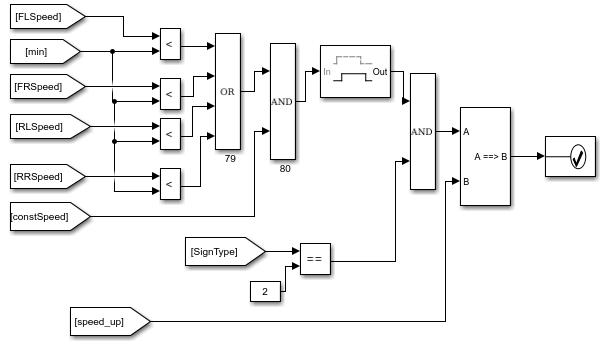}
  \caption{Model of requirement R10: If the vehicle detects minimum speed limits sign when \emph{constSpeed} state is active, and its speed is less than the speed limit (\emph{SignType} equals 2), \emph{speed\_up} state will be activated in the next time step.}
  \label{R10}
\end{figure}

\begin{figure}[htbp]
  \centering
  \includegraphics[width=5.5in]{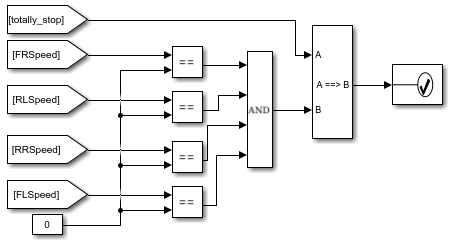}
  \caption{Model of requirement R25: When \emph{totally\_stop} state is active, the speed of the four wheels of the vehicle should be zero.}
  \label{R25}
\end{figure}

The non-functional properties includes {\gt{energy}} properties and temporal properties. SDV is not applicable for verifying the timing constraints and {\gt{energy}} properties because it lacks descriptive blocks for constructing the requirements. For constructing the timing constraints, to detect the time instant of the input arrival or the output departure, the trigger-based blocks should be applied. However, SDV does not support the trigger-based blocks. Moreover, when modeling the behaviour of {\gt{energy}} consumption, we add the stochastic elements (blocks/algorithms), which is not supported in SDV, either.

Table \ref{table_verification_result} (where the variables with prime represent variables in the previous time step) shows the verification result of the requirements. For the functional properties that are not related to time (R1 to R25, R30 and R31), the verification results are all valid in SDV. SDV cannot provide the verification results (valid or not) of R32 to R36 because the verification time exceeds the user-defined maximum analysis time (here it is set as 50 minutes).

\section{\uppaal Verification and Simulation}
A validation of the translation patterns becomes a reachability analysis in the following forms:
\begin{inparaenum}
\item $A[\ ]$ $\neg$$deadlock$ verifies that a system is free of any inconsistencies iff
there is no deadlock and all the constraints modeled in STA through the mapping strategy are satisfied;
\item $A[\ ]$ $\neg$$(STA.fail)$, where $fail$ is a location containing
constraints that are not satisfied. It verifies that a given constraint modeled with an STA  never
reaches the $fail$ location. 
\end{inparaenum}

\begin{figure}[htbp]
  \centering
  \includegraphics[width=5.5in]{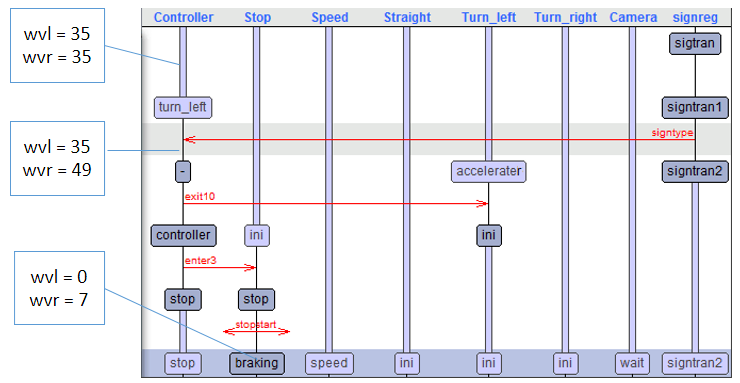}
  \caption{CE in \smc: Initially, { \gt{Controller}} transits to \emph{turn\_left}. When { \gt{Turn\_left}} stays in \emph{accelerater} location, the \emph{wvr} is increased while \emph{wvl} is constant. When the vehicle detects a stop sign and brakes to stop, finally \emph{wvl} is 0 but \emph{wvr} is larger than 0 hence violates R16.}
  \label{CE}
\end{figure}

\begin{figure}[htbp]
\centering
\subfigure[signType and stop mode]{
\label{CE1}
\includegraphics[width=3in]{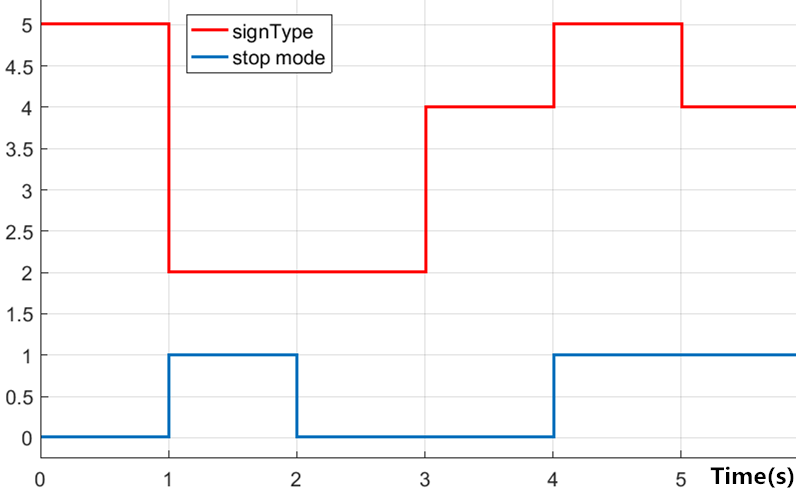}}
\subfigure[speeds of the left/right wheels]{
\label{CE2}
\includegraphics[width=3in]{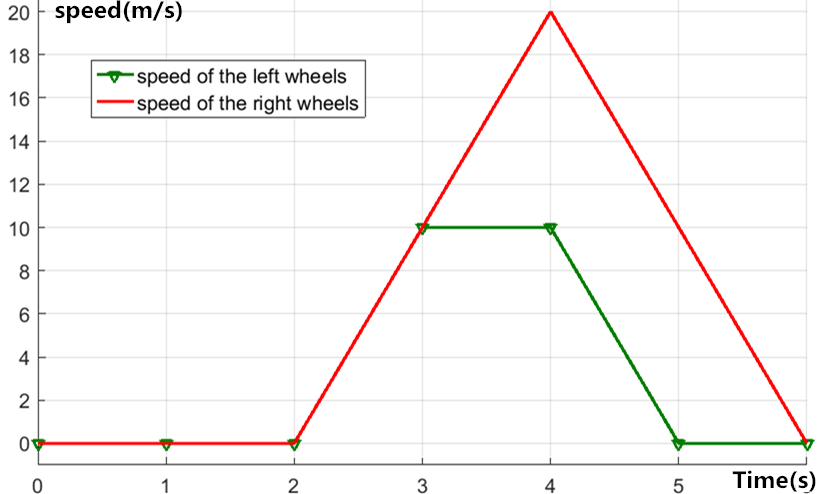}}
\caption{CE in SDV: signType is turn left (4) at 3s and becomes stop (5) at 4s. Then the vehicle enters \emph{stop} mode and decreases speeds of the left and right wheels. However, at 5s, the speed of the left wheels is 0 while the speed of the right wheels is greater than 0, which disproves R16.}
\label{SDVCE}
\end{figure}

When verifying R16, it is proved that R16 is invalid in both SDV and \smc. By tracing the counter-example illustrated in Fig.\ref{CE} and Fig.\ref{SDVCE}, we can find the reason of the invalidity: If the vehicle detects a stop sign when turning left/right, it will enter stop state with the speeds of the left/right wheels unequal. Since the acceleration of the wheels are identical, the speed will not be decreased to zero simultaneously.
We refine the requirement of R16 as: when the vehicle detects a stop sign when turning left/right, it should first finish the turning and then stop. The requirement becomes valid after modifying the model.




\begin{sidewaystable}[htbp]
\begin{longtable}[htbp]{|c|c|p{325pt}|c|c|c|c|}
  \caption{\large{Logic expression and verification results}}\\ \hline
    Req & Type & Expression & Result & Time (min) & Memory (Mb) & CPU (\%))\\
    \hline
    \multirow{2}{*}{R1} & SDV & ${\mathop{constspeed'}==true \wedge signType==4 \implies  turn\_left==true }\  $ & valid & 6 & 1834 & 40.17 \\\cline{2-7}
    & \smc & ${Pr[\leqslant 3000]([\ ] (CONSTSPEED==1 \wedge signType==4 }  {\wedge \neg Ctrl.ctrl} $\newline$ \wedge Ctrl.t==0)     {\implies } { Ctrl.turn\_left) \geqslant 0.95 }$ \footnote{Experts in TCTL will recognize that $\varphi \leadsto \psi$ is equivalent to $A[\ ] (\varphi \implies A <> \psi)$} & valid & 0.002 & 30.91 & 1.2  \\
    \hline

    \multirow{2}{*}{R2} & SDV &  ${\mathop{speed\_up'} == true \wedge signType == 4 \implies turn\_left == true}$ & valid & 6.25 & 1855 & 39.61 \\\cline{2-7}
    & \smc & ${Pr[\leqslant 3000]([\ ] (SPEEDUP==1 \wedge signType==4 }   \wedge \neg Ctrl.ctrl $\newline$ \wedge Ctrl.t==0) \implies  { Ctrl.turn\_left) \geqslant 0.95 }$ & valid & 0.002 & 30.91& 1.2\\
    \hline

    \multirow{2}{*}{R3} & SDV & ${\mathop{speed\_down'} ==true \wedge signType == 4 \implies turn\_left == true }$& valid & 6.25 & 1849 & 39.30\\\cline{2-7}
    & \smc & ${Pr[\leqslant 3000]([\ ] (SPEEDDOWN==1 \wedge signType==4  \wedge}\neg Ctrl.ctrl $\newline$ \wedge Ctrl.t==0)  {\implies }{ Ctrl.turn\_left) \geqslant 0.95 }$ & valid & 0.002 & 30.91 & 1.2 \\
    \hline

    \multirow{2}{*}{R4} & SDV &${ \mathop{constspeed'} == true \wedge signType == 3 \implies turn\_right == true}$  & valid & 6.30 & 1872 & 39.42\\\cline{2-7}
    & \smc & ${Pr[\leqslant 3000]([\ ] (CONSTSPEED==1 \wedge signType==3 }  { \wedge \neg Ctrl.ctrl} $\newline$ \wedge Ctrl.t==0) \implies  { Ctrl.turn\_right) \geqslant0.95}$ & valid & 0.003 & 31.02 & 1.3  \\
    \hline

    \multirow{3}{*}{R5} & SDV & ${\mathop{speed\_up'} == true \wedge signType == 3 \implies turn\_right == true}$ & valid & 6.46 & 1872 & 40.14\\\cline{2-7}
    & \smc & ${Pr[\leqslant 3000]([\ ] (SPEEDUP==1 \wedge signType==3  }  {\wedge \neg Ctrl.ctrl }$\newline$\wedge Ctrl.t==0) \implies { Ctrl.turn\_right)\geqslant0.95}$ & valid & 0.003 & 31.02 & 1.3 \\
    \hline

    \multirow{3}{*}{R6} & SDV & ${\mathop{speed\_down'} ==true \wedge signType == 3 \implies turn\_right == true}$ & valid & 7.04 & 1895 & 40.16\\\cline{2-7}
    & \smc & ${Pr[\leqslant 3000]([\ ] (SPEEDDOWN==1 \wedge signType==3 \wedge  }  \neg Ctrl.ctrl$\newline$ \wedge Ctrl.t==0) \implies { Ctrl.turn\_right)\geqslant 0.95}$ & valid & 0.003 & 31.02 & 1.3 \\
    \hline

    \multirow{3}{*}{R7} & SDV & ${\mathop{constspeed'} == true \wedge signType == 5 \implies braking == true}$ & valid & 8.47 & 1953 & 39.61\\\cline{2-7}
    & \smc & ${Pr[\leqslant 3000]([\ ] (signType==5 \wedge CONSTSPEED==1 \wedge }   \neg Ctrl.ctrl $\newline$\wedge { Ctrl.t==0) \implies Ctrl.stop)\geqslant 0.95}$ & valid & 0.003 & 30.91 & 1.3 \\
    \hline

    \multirow{3}{*}{R8} & SDV & ${\mathop{speed\_up'} == true \wedge signType == 5 \implies braking == true}$ & valid & 8.47 & 1953 & 39.61\\\cline{2-7}
    & \smc & ${Pr[\leqslant 3000]([\ ] (signType==5 \wedge SPEEDUP==1 \wedge  }  \neg Ctrl.ctrl $\newline$ \wedge { Ctrl.t==0) \implies Ctrl.stop)\geqslant 0.95}$ & valid & 0.003 & 30.91 & 1.3 \\
     \hline

    \multirow{3}{*}{R9} & SDV & ${\mathop{speed\_down'} == true \wedge signType == 5 \implies braking == true}$ & valid & 6.36 & 1905 & 39.39\\\cline{2-7}
    & \smc & $ Pr[\leqslant 3000]([\ ] (signType==5 \wedge SPEEDDOWN==1 \wedge \neg Ctrl.ctrl $\newline$ \wedge{ Ctrl.t==0) \implies Ctrl.stop)\geqslant 0.95}$ & valid & 0.003 & 30.91 & 1.3 \\
    \hline
  \label{table_verification_result}%
  \end{longtable}
\end{sidewaystable}

\begin{sidewaystable}[htbp]

\begin{longtable}[htbp]{|c|c|p{315pt}|c|c|c|c|}
   \hline
    Req & Type & Expression & Result & Time (min) & Memory (Mb) & CPU (\%))\\
    \hline
    \multirow{2}{*}{R10} & SDV & $constspeed' == true \wedge signType == 2 { \wedge (FLS < min \vee FRS < min }  \vee RLS < min \vee RRS < min) \implies {speed\_up == true}$ & valid & 7.18 & 1894& 39.28\\\cline{2-7}
    & \smc& ${Pr[\leqslant 3000]([\ ] (Straight.CONSTSPEED==1 \wedge sign==2  }  \wedge (FLS<speedl \vee FRS<speedl \vee RLS<speedl \vee RRS<speedl) \wedge Straight.t==0) \implies Straight.speed\_up)\geqslant0.95 $ & valid & 0.003 & 30.92 & 1.4 \\
    \hline
    \multirow{3}{*}{R11} & SDV & ${constspeed'} == true \wedge signType == 1 \wedge (FLS > max \vee FRS > max \vee RLS > max \vee RRS > max)  \implies  speed\_down == true$ & valid & 6.40 & 1931 &39.43\\\cline{2-7}
    & \smc& $Pr[\leqslant 3000]([\ ] (Straight.CONSTSPEED==1 \wedge sign==1 \wedge (FLS>speedh \vee FRS>speedh \vee RLS>speedh $ $\vee RRS>speedh) \wedge Straight.t==0) \implies Straight.speed\_down)\geqslant 0.95 $ & valid & 0.002 & 30.92 & 1.2\\
    \hline

    \multirow{3}{*}{R12} & SDV & $\mathop{speed\_up'} == true \wedge signType == 1  \wedge (FLS > max \vee FRS > max \vee RLS > max \vee RRS > max) \implies speed\_down $ & valid & 11.58 & 1977 &37.41\\\cline{2-7}
    & \smc& $Pr[\leqslant 3000]([\ ] (Straight.UP==1 \wedge sign==1    \wedge (FLS>speedh \vee FRS>speedh \vee RLS>speedh  \vee RRS>speedh) \wedge Straight.t==0) \implies Straight.speed\_down)$ & valid & 0.002 & 30.95 & 22.24\\
    \hline

    {R13} & SDV & $constspeed' == true \wedge signType == 2    \wedge FLS \geqslant min \geqslant FRS \geqslant min    \wedge RLS \geqslant min \wedge RRS \geqslant min \implies constspeed $ & valid & 5.37 & 2085 & 39.05\\
    \hline

    \multirow{3}{*}{R14} & SDV & $speed\_down' == true \wedge signType == 2 \wedge(FLS < min \vee FRS < min   \vee RLS < min \vee RRS < min) \implies speed\_up$ & valid &10.21& 1947& 38.93\\\cline{2-7}
    & \smc& $Pr[\leqslant 3000]([\ ] (Straight.DOWN==1 \wedge sign==2   \wedge (FLS<speedl \vee FRS<speedl \vee RLS<speedl  \vee RRS<speedl) \wedge Straight.t==0) \implies Straight.speed\_up)\geqslant 0.95$ & valid & 0.002 & 30.95 & 1.1\\
    \hline
	
 {R15} & SDV & $constspeed' == true \wedge signType == 1 \wedge FLS \leqslant max \wedge FRS \leqslant max  \wedge RLS \leqslant max \wedge RRS \leqslant max \implies constspeed $ & valid & 5.10& 2095& 40.20\\\cline{2-7}
    \hline

    \multirow{3}{*}{R16} & SDV & $turn\_left' ==true \wedge signType == 5 \implies stop\_flag ==0 $ & valid & 0.15& 2038& 38.76\\\cline{2-7}
    & \smc& $Pr[\leqslant3000]([\ ] (LEFT==1 \wedge sign==5 \wedge \neg Controller.controller)   \implies Controller.stop)\geqslant0.95 $ & valid & 1.31 & 30.97 & 2.8\\
    \hline

    \multirow{1}{*}{R17} & SDV & $turn\_right' ==true \wedge signType == 5 \implies stop\_flag ==0 $  & valid & 0.15& 1963& 23.04\\ \cline{2-7}
    & \smc& $Pr[\leqslant3000]([\ ] (RIGHT==1 \wedge sign==5 \wedge \neg \  Controller.controller)   \implies Controller.stop)\geqslant0.95 $& valid & 1.2 & 30.98 & 2.6\\
    \hline

    {R18} & SDV & $speed\_up' == true \wedge signType == 0 \implies speed\_up$ & valid & 11.52& 2026& 39.99\\
    \hline

    {R19} & SDV & $speed\_down' ==  true \wedge signType == 0 \implies speed\_down$&valid & 11.13 & 2054& 40.09\\
    \hline
  \label{table_logic2}%
  \end{longtable}
\end{sidewaystable}

\begin{sidewaystable}[htbp]

\begin{longtable}[htbp]{|c|c|p{315pt}|c|c|c|c|}
   \hline
    Req & Type & Expression & Result & Time (min) & Memory (Mb) & CPU (\%))\\
    \hline
    \multirow{3}{*}{R20} & SDV & $turn\_left' == true \wedge FLS \geqslant 70 \wedge FRS \geqslant 70   \wedge RLS \geqslant 70 \wedge RRS \geqslant 70 \wedge turn\_left == true \implies TRDL == true$ & valid & 0.04 & 1961& 4.38\\\cline{2-7}
    & \smc& $Pr[\leqslant3000]([\ ] (Controller.turn\_left \wedge entry\_speed\geqslant70)   \implies (Turn\_left.deceleratel || Turn\_left.decelerater)) $ & valid & 0.002 & 34.07 & 1.2\\

    \multirow{3}{*}{R21} & SDV & $turn\_left' == true \wedge FLS < 70 \wedge FRS < 70 \wedge RLS < 70 \wedge RRS < 70 \wedge turn\_left == true \implies TRAR == true $& valid & 0.01 & 1961& 4.38\\\cline{2-7}
    & \smc& $Pr[\leqslant3000]([\ ] (Controller.turn\_left \wedge entry\_speed<70) \implies (Turn_left.accelerater || Turn\_left.acceleratel)) $ &valid & 0.002 & 34.07 & 1.3\\
    \hline

    \multirow{3}{*}{R22} & SDV & $turn\_right' == true \wedge FLS \geqslant 70 \wedge  FRS \geqslant 70  \wedge RLS \geqslant 70 \wedge RRS \geqslant 70 \wedge turn\_right == true \implies TRDR == true $ & valid &0.03 & 2234& 3.92\\\cline{2-7}
    & \smc& $Pr[\leqslant3000]([\ ] (Controller.turn\_right \wedge entry\_speed\geqslant70) \implies (Turn\_right.deceleratel || Turn\_right.decelerater)) $ & valid & 0.002 & 34.07 & 1.1\\
    \hline

    \multirow{3}{*}{R23} & SDV & $turn\_right' == true \wedge FLS < 70 \wedge FRS < 70 \wedge RLS < 70 \wedge RRS < 70 \wedge turn\_right == true \implies TRAL == true$ & valid &0.01 & 1965& 8.36\\\cline{2-7}
    & \smc& $Pr[\leqslant3000]([\ ] (Controller.turn\_right \wedge entry\_speed<70) \implies (Turn\_right.accelerater || Turn\_right.acceleratel))$ & valid & 1.08 & 34.07 & 2.5\\
    \hline

    \multirow{3}{*}{R24} & SDV & $braking' == true \wedge braking == false \implies FLS ==0 \wedge FRS ==0 \wedge RLS ==0 \wedge RRS == 0$  & valid & 1.10 & 1918&  38.48\\\cline{2-7}
    & \smc& $Pr[\leqslant3000]([\ ] Stop.braking \implies (FLS==FRS\wedge RLS==RRS \wedge FLS==RLS)) $ & valid & 4 & 34.89 & 4.1\\
    \hline

    \multirow{3}{*}{R25} & SDV & $totally\_stop == true \implies FLS ==0 \wedge FRS ==0 \wedge RLS ==0 \wedge RRS == 0$   & valid & 0.31 & 1930&  24.02\\\cline{2-7}
    & \smc& $Pr[\leqslant3000]([\ ] Stop.totally_stop \implies (FLS==FRS\wedge RLS==RRS\wedge FLS==RLS)) $ & valid & 0.003 & 30.96 & 1.6\\
    \hline

    {R26}  & \smc& $Pr[\leqslant3000]([\ ] speedh == 100 \implies FLS \leqslant 90) \geqslant Pr[\leqslant3000]([\ ] speedh == 100 \implies FLS > 90 \wedge FLS \leqslant 100)$ & valid & 17.3 & 29.28 & 13.2\\
    \hline

    {R27}  & \smc& $Pr[\leqslant3000]([\ ] speedh == 120 \implies FLS \leqslant 110) \geqslant Pr[\leqslant3000]([\ ] speedh == 120 \implies FLS > 110 \wedge FLS \leqslant 120)$ & valid & 0.22 & 30.06 & 1.5\\
    \hline
    {R28}  & \smc& $Pr[\leqslant3000]([\ ] speedl == 70 \implies FLS \geqslant 80) \geqslant Pr[\leqslant3000]([\ ] speedl == 70 \implies FSL > 70 \wedge FLS \leqslant 80)$  &invalid & 2.45 & 28.55 & 11.2\\
    \hline

    {R29}  & \smc& $Pr[\leqslant3000]([\ ] speedl == 80 \implies FLS \geqslant 90) \geqslant Pr[\leqslant3000]([\ ] speedl == 80 \implies FLS > 80 \wedge FLS \leqslant 90)$ & valid & 21.85 & 29.57 & 16.3\\
    \hline

  \label{table_logic2}%
  \end{longtable}
\end{sidewaystable}

\begin{sidewaystable}[htbp]
\begin{longtable}[htbp]{|c|c|p{315pt}|c|c|c|c|}
   \hline
    Req & Type & Expression & Result & Time (min) & Memory (Mb) & CPU (\%))\\
    \hline

   \multirow{3}{*}{R30} & SDV & $turn\_left == true \implies FLS \leqslant FRS \wedge RLS \leqslant RRS $  & valid &31.26 & 2253&  38.86\\\cline{2-7}
    & \smc& $Pr[\leqslant3000]([\ ] Ctrl.turn\_left \implies (FLS\leqslant FRS)) $ & valid & 0.001 & 30.96 & 0.9\\
    \hline

    \multirow{3}{*}{R31} & SDV & $turn\_left == true \implies FLS \leqslant FRS \wedge RLS \leqslant RRS $  & valid &49.15  & 2253&  38.85\\\cline{2-7}
    & \smc& $Pr[\leqslant3000]([\ ] Controller.turn\_right \implies (FLS\geqslant FRS)) $ & valid & 0.002 & 30.96 & 1.2\\
    \hline
    \multirow{3}{*}{R32} & SDV & $speed\_up' == true \wedge speed\_up ==false \implies speed\_up\_time < 2400$  & undecided &59.15  & 2125&  36.40\\\cline{2-7}
    & \smc& $Pr[\leqslant3000]([\ ] Straight.speed\_up \implies Straight.t\leqslant120) $ & undecided & 23:03 & 1167.54 & 22.24\\
    \hline
        \multirow{3}{*}{R33} & SDV & $speed\_down'== true \wedge speed\_down ==false \implies speed\_down\_time < 2400$  & undecided &60.15  & 2282&  36.8\\\cline{2-7}
    & \smc& $Pr[\leqslant3000]([\ ] Straight.speed_down \implies Straight.t\leqslant120)$  & valid & 0.002 & 30.96 & 1.2\\
    \hline

    \multirow{3}{*}{R34} & SDV & $braking == true \wedge braking ==false \implies braking\_time < 2400 $& undecided &49  & 2282&  37.11\\\cline{2-7}
    & \smc & $Pr[\leqslant3000]([\ ] \neg BrakeExec.fail)\geqslant0.95$ & valid & 0.002 & 30.964 & 1.4\\
    \hline

    \multirow{3}{*}{R35} & SDV & $turn\_left' == true \wedge turn\_left == false \implies turn\_left\_time \geqslant 800 \wedge turn\_left\_time \leqslant 1500$ & undecided &60.05  & 2282&  37.66\\\cline{2-7}
    & \smc & $Pr[\leqslant3000]([\ ] Ctrl.turn\_left \wedge \neg Turn\_left.ini \implies Ctrl.left\_clk\leqslant 75)\geqslant0.95 $ & valid & 0.002 & 30.96 & 1.1\\
    \hline

    \multirow{3}{*}{R36} & SDV & $turn\_right' == true \wedge turn\_right == false \implies turn\_right\_time \geqslant 800 \wedge turn\_right\_time \leqslant 1500 $  & undecided &55.47  & 2282&  37.72 \\\cline{2-7}
    & \smc & $Pr[\leqslant3000]([\ ] Ctrl.turn\_right \wedge \neg Turn\_right.ini \implies Ctrl.right\_clk\leqslant 75)\geqslant0.95 $&valid & 0.003 & 30.96 & 1.6\\
    \hline
    \multirow{3}{*}{R37} & \smc& $Pr[\leqslant3000]([\ ] Camera.CamExec_en \leqslant 3)\geqslant0.95$& valid & 12.4  & 32.91 & 22 \\\cline{2-7}
    & \smc & $simulate1[\leqslant3000]\{camera\_en\}$& valid &0.22  & 31.65&  2.6\\
    \hline

    \multirow{3}{*}{R38} & \smc& ${Pr[\leqslant3000]([\ ] signreg.RegExec_en \leqslant 5)\geqslant0.95}$& valid &12.9  & 32.9&  24.5\\\cline{2-7}
    & \smc & ${simulate1[\leqslant3000]\{signreg\_en\}}$& valid &0.23  & 31.62&  1.8\\
    \hline

    {R39} & \smc& ${simulate1[\leqslant3000]\{signType, average\_speed, constSpeed\_en\}}$& valid &0.23  & 31.57&  2.5\\
    \hline
\multirow{3}{*}{R40} & \smc& ${Pr[\leqslant3000]([\ ] Turnning\_en \leqslant 270)\geqslant0.95}$& valid & 13.46  & 30.24&  23.6\\\cline{2-7}
    & \smc& ${simulate1[\leqslant3000]\{signType, average\_speed, Turning\_en\}}$& valid & 0.22  & 31.63&  2.8\\
    \hline

    \multirow{3}{*}{R41} & \smc& ${Pr[\leqslant3000]([\ ] Turnning\_en \leqslant 270)\geqslant0.95}$& valid & 13.46  & 30.24&  23.6\\\cline{2-7}
    & \smc& ${simulate1[\leqslant3000]\{signType, average\_speed, Turning\_en\}}$& valid & 0.22  & 31.63&  2.8\\
    \hline

    \multirow{2}{*}{R42} &\smc~ & ${E[\geqslant3000;100]([\ ] ~ max:energy.braking\_en)}$ & valid & 32.53 & 36.51 & 15.31\\\cline{2-7}
    & \smc& ${simulate1[\leqslant3000]\{signType, average\_speed, braking\_en\}}$& valid & 0.002 & 30.96 & 1.2\\
    \hline

    \multirow{3}{*}{R43} & \smc& ${Pr[\leqslant3000]([\ ] Up\_en \leqslant 400)\geqslant0.95}$& valid &13.39  &30.46 &  21.6\\\cline{2-7}
    & \smc& ${simulate1[\leqslant3000]\{signType, average\_speed, Up\_en\}}$& valid &0.22  & 31.63&  2.3\\
    \hline
  \label{table_logic3}%
  \end{longtable}
\end{sidewaystable}

\begin{sidewaystable}[htbp]
\begin{longtable}[htbp]{|c|c|p{315pt}|c|c|c|c|}
   \hline
    Req & Type & Expression & Result & Time (min) & Memory (Mb) & CPU (\%))\\
    \hline
    \multirow{3}{*}{R44} & \smc& ${Pr[\leqslant3000]([\ ] Down\_en \leqslant 400)\geqslant0.95}$& valid &13.63  & 30.24&  19.8\\\cline{2-7}
    & \smc& ${simulate1[\leqslant3000]\{signType, average\_speed, Down\_en\}}$& valid &0.24  & 31.63&  2.5\\
    \hline

    {R45} & \smc& ${simulate1[\leqslant3000]\{average\_speed, Con\_en\}}$& valid &0.23  & 31.57&  2.8\\
    \hline

    \multirow{2}{*}{R46} & \smc& ${Pr[\leqslant3000]([\ ]\neg SignRegExec.fail) \geqslant 0.95}$& valid & 0.02  & 30.96&  1.6\\\cline{2-7}
    & \smc& ${simulate100[\leqslant3000]\{signType, cameraexec\}}$& valid & 0.19 & 31.01 & 2.3\\
    \hline

    \multirow{2}{*}{R47} & \smc& ${Pr[\leqslant3000]([\ ] \neg CameraExec.fail) \geqslant 0.95}$& valid &0.01  & 30.96 &  1.3\\\cline{2-7}
    & \smc& ${simulate100[\leqslant3000]\{cameraExec\}}$ & valid & 0.18 & 30.80 & 4.0\\
    \hline

    \multirow{2}{*}{R48} & \smc& ${Pr[\leqslant3000]([\ ]\neg Synchronization.fail) \geqslant 0.95}$& valid &0.02   & 30.96&  2.1\\\cline{2-7}
    & \smc& ${simulate100[\leqslant3000]\{signType, s_1, s_2, s_3, s_4\}}$ & valid & 1.5 & 56.29 & 20.49\\
    \hline

    \multirow{2}{*}{R49} & \smc& $Pr[\leq3000]([\ ] \neg Periodic.fail) \geqslant 0.95$& valid &0.02  & 30.96&  1.8\\\cline{2-7}
    & \smc& ${simulate100[\leqslant3000]\{cameraexec\}}$& valid & 0.18 & 30.80 & 4.0\\
    \hline

    \multirow{2}{*}{R50} & \smc& $Pr[\leqslant3000]([\ ]\neg End-to-End.fail) \geqslant 0.95$ & valid &1.10   & 30.96&  17.1\\\cline{2-7}
    & \smc& $simulate100[ \leqslant 3000]\{camera.Camera,signreg.SignReg\}$& valid & 0.04 & 37.80 & 2.22\\
    \hline

    {R51} & \smc& $Pr[\leqslant 3000]$({ $[\ ]$}$ CamToReg.dclk \geqslant worst\_camexec + worst\_signregexec)$ & [0.9,1] &  11.83 & 29.62 & 11.2\\
    \hline
  \label{table_logic3}%
  \end{longtable}
\end{sidewaystable}

The verification results using \smc\ are established as valid with 95\% confidence and given in Table \ref{table_verification_result}.  The time bound on the simulations is set to 3000 time units (60s) and covers most cases of {\gt{signType}} sequences. We also run one hundred simulations for each requirement separately. Regular \uppaal~can not provide the verification results (valid or not) of non-functional requirements because the verification consumes long time and memory. Table \ref{table_verification_result} presents the verification results of all properties in \smc. When verifying the temporal properties with large time steps, it takes a long time. To solve this problem, the time was scaled down in \smc. Most functional properties (R1 to R25, R30 to R36) and energy- (R37 to R41, R43 to R45) and timing constraint requirements (R46 to R51) are valid with probability 0.95.

To guarantee safety of our \av, we verify R26 that when our car detects a maximum speed limit sign, even if the speeds of the car is not so high but satisfies maximum speed limit, the car will not increase its speed. The probability that the speeds of the car is less than 90 should be larger than the probability that the speeds of the car is in [90,100] is accepted with 95\% confidence. Similar queries (R27 to R29) are provided for maximum speed limit to keep safety of the vehicle system.

\begin{figure}[htbp]
\centering
\subfigure[Histogram of probability distribution of energy estimation]{
\label{fig:R7_histo}
\includegraphics[width=3.5in]{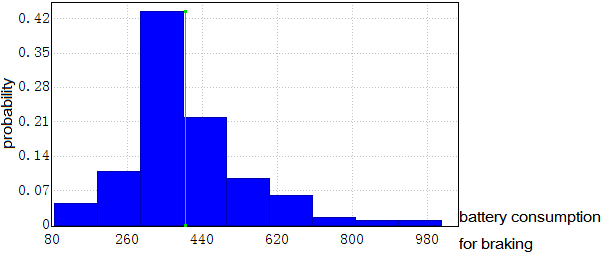}}
\subfigure[Pie chart of probability]{
\label{fig:R7_pie_chart}
\includegraphics[width=2.8in]{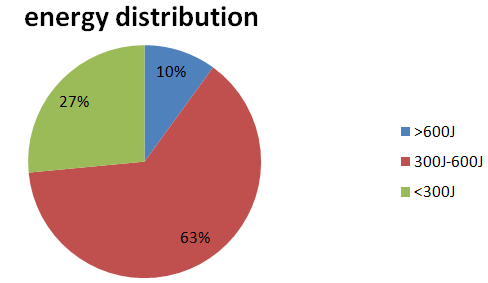}}
\caption{Energy estimation of braking mode}
\label{energy_result_prob}
\end{figure}

The frequency histogram shown in Fig.\ref{fig:R7_histo} is the result of query for R42 in Table \ref{table_verification_result}. \smc~evaluates the maximum battery consumption for braking the car within 3000 time units (60s) for 100 runs and generates frequency histogram. According to the graph, the average energy consumption is approx. 400J (green line). As shown in \ref{fig:R7_pie_chart}, the energy consumption of braking the car is more likely to be between 300J to 600J (with probability 63\%).

The probability that the {\gt{End-to-End}} time from inputs of {\gt{Camera}} to outputs of {\gt{Sign Recognition}} is less than or equal to sum of worst execution time of the two \fp s (R51) is provided within [0.9, 1] with 95\% level of significance. \emph{worst\_camexec} and \emph{worst\_signregexec} represents for worst execution time of {\gt{Camera}} and {\gt{Sign Recognition}} respectively, which are 100ms and 400ms according to our requirements.

\section{\simu~\& \staf~ Simulation using \mt~ tool}

In our experiments, simulation is conducted in Simulink and \smc~to validate both functional properties and the non-functional properties.

\begin{figure}[ht!]
  \centering
  \includegraphics[width=4in]{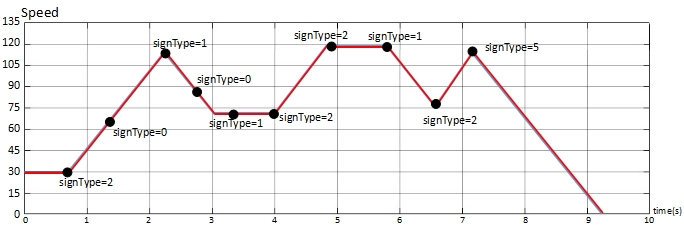}
  \caption{The simulation result of maximum and minimum speed limit in S/S model}
  \label{fig:LandU}
\end{figure}
\begin{figure}[ht!]
  \centering
  \includegraphics[width=4in]{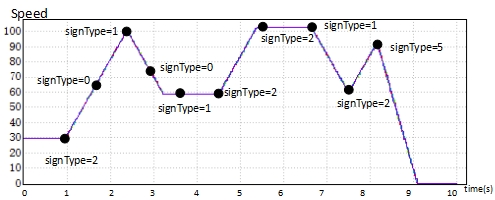}
  \caption{The simulation result of maximum and minimum speed limit in \smc~model}
  \label{fig:LandUu}
\end{figure}

In this chapter, we present several simulation results of the requirements on the vehicle. For example, as shown in Fig.\ref{fig:LandU}, the requirements of R8, R10-R15, R18, R19, R24, R25, and R32-R34 are validated.
When \emph{time} equals to 0.61s and a \emph{minimum\ speed\ limit} sign (\emph{signType} == 2) is detected, since the current speed of the vehicle is 30m/s (less than \emph{min\_speed\_limit} 70), the speed of the vehicle is increased. Thus R10 is validated.
During the acceleration, a \emph{straight} sign is recognized, and the vehicle keeps accelerating, which satisfies R18.
At 2.2s, the vehicle recognizes a \emph{maximum\ speed\ limit} sign (signType == 1 and the speed\_limit is 100), it then decreases its speed. R14 is then guaranteed.
Afterwards, the vehicle detects a \emph{straight} sign, it continues decelerating its speed. In this case, R19 is satisfied.
After completing the deceleration, the vehicle maintains the speed. At 3.4s, a \emph{maximum\ speed\ limit} sign (100) is recognised, because the current speed of the vehicle is less than 100 according to R15, the speed of the vehicle is unchanged until 4s, at which the vehicle detects a \emph{minimum\ speed\ limit} sign (80). The vehicle increases its speed to 119m/s.
Then a \emph{minimum\ speed\ limit} sign appears again, but as the speed of the vehicle now is 119, larger than \emph{min\_speed\_limit} 70, the vehicle will keep its speed as R13.
At 5.76s, a \emph{maximum\ speed\ limit} sign is recognised, and the vehicle decreases its speed to satisfy the \emph{max\_speed\_limit} which is 100, which satisfies R11.
During the deceleration, if a \emph{minimum\ speed\ limit} sign 80 is recognised and the speed of the vehicle is lower than 80, the vehicle will increase its speed according to R12.
Finally, at 7.1s during acceleration, a \emph{stop} sign is recognised, so the vehicle decreases the speed of its four wheels till 0, which validates R8.
According to Fig.\ref{fig:LandU}, the longest acceleration time is 1.59s and the longest deceleration time is 0.82s, thus R32 and R33 is proved. The deceleration time for stop is 2.06s, satisfying R34.
The differences between the speeds of the (front and rear) wheels on the left and right are nearly 0, as the four lines representing speed are almost overlapped. Similarly, the speeds of the four wheels are shown, and the requirements mentioned above can be proved from Fig.\ref{fig:LandUu}.

\begin{figure}[ht!]
  \centering
  \includegraphics[width=4in]{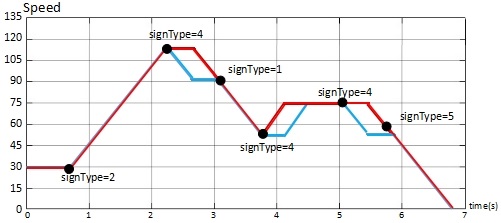}
  \caption{The simulation result of turning left in S/S model}
  \label{turningl}
\end{figure}
\begin{figure}[ht!]
  \centering
  \includegraphics[width=4in]{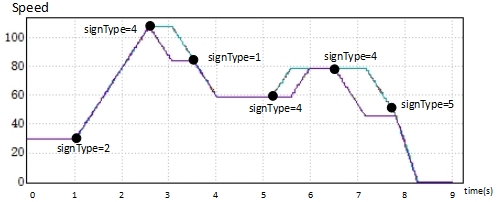}
  \caption{The simulation result of turning left in \smc~model}
  \label{turninglu}
\end{figure}

In Fig.\ref{turningl}, R1-3, R16, R20-R21, R30 and R35 can be validated.
At 0.64s, a \emph{minimum\ speed\ limit} sign (signType == 2) is recognised, and as the speed of the vehicle now is 30m/s, which is less than the \emph{min\_speed\_limit} 80, so the vehicle will increase its speed.
During the acceleration, at 2.23s, a \emph{turn\_left} sign (signType == 4) is recognised. As the speed of the vehicle now is 112, larger than 70 (high speed), so the vehicle will decrease the speeds of rear and front left wheels to turn left, which satisfies R2 and R20.
At 3.78s, as the speed of the vehicle is 52, less than 70 (low speed), the vehicle will turn left as a \emph{turn\_left} sign is recognised. This time, the turning is realized by increasing speeds of rear and front right wheels, according to R3 and R21.
After the vehicle turns left, the vehicle will run at a constant speed, when at 5.05s, a \emph{turn\_left} sign is recognised. So the vehicle will turn left, which validates R1.
During the turning, at 5.69s, a \emph{stop} sign is recognised. The vehicle will finish turning, and then decrease to stop. Thus R16 is validated. Finally, the vehicle will stop with all the speed of the four wheels equals to 0, thus validates R24.
The turning time are 0.906s, 0.8s and 0.91s, within 0.8s to 1.5s, thus R35 is validated.
Similarly, the behaviour of the vehicle's turning can be seen from Fig.\ref{turninglu}, and the requirements can be validated from differences among speeds. The vehicle will turn left by either increasing speed of the right wheels or decreasing speed of the left wheels depends on the speed of the vehicle, and the turning time is within [0.8s,1.5s], thus the requirements R30 and R35 can be valid.

\begin{figure}[ht!]
  \centering
  \includegraphics[width=4in]{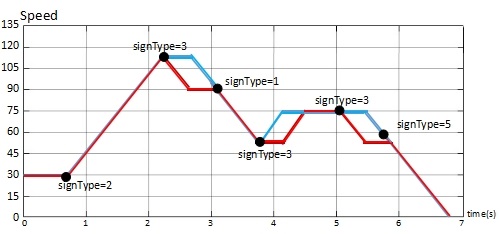}
  \caption{The simulation result of turning right in S/S model}
  \label{turningr}
\end{figure}
\begin{figure}[ht!]
  \centering
  \includegraphics[width=4in]{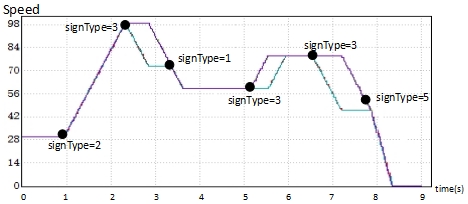}
  \caption{The simulation result of turning right in \smc ~model}
  \label{turningru}
\end{figure}

In Fig.\ref{turningr}, R4-6, R17, R22-R23, R31 and R36 can be validated.
At 0.62s, a \emph{minimum\ speed\ limit} sign (signType == 2) is recognised, and as the speed of the vehicle now is 30m/s, which is less than the \emph{min\_speed\_limit} 70, so the vehicle will increase its speed.
During the acceleration, at 2.2s, a \emph{turn\_right} sign (signType == 3) is recognised. As the speed of the vehicle now is 110, larger than 70 (high speed), so the vehicle will decrease the speeds of rear and front right wheels to turn right, which satisfies R5 and R22.
At 3.76s, when the vehicle is decelerating, it recognised a turn right sign. Thus it turn left, which satisfies R6.
As the speed of the vehicle is 55, less than 70 (low speed), the vehicle will turn right as a \emph{turn\_right} sign is recognised. This time, the turning is realized by increasing speeds of rear and front left wheels according to R23.
After the vehicle turns right, the vehicle will run at a constant speed, when at 5.0s, a \emph{turn\_right} sign is recognised. So the vehicle will turn right according to R4.
During the turning, at 5.72s, a \emph{stop} sign is recognised. The vehicle will finish turning, and then decrease to stop. Thus R17 can be validated. Finally, the vehicle will stop with all the speed of the four wheels equals to 0, thus validates R24. The turning time are 0.91s, 0.82s and 0.915s, within 0.8s to 1.5s, thus R36 is validated. During the whole run of the vehicle, when it is turning right, the speed of the left wheels are always no less then that of the right wheels. Thus R31 is valid. Recall Fig.\ref{turningru}, same requirements can be valid in \smc~model.

\begin{figure}[ht!]
  \centering
  \includegraphics[width=4in]{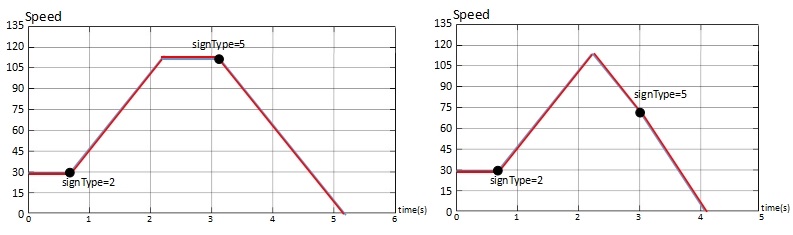}
  \caption{The simulation result of stop}
  \label{stopsimu}
\end{figure}
\begin{figure}[ht!]
  \centering
  \includegraphics[width=4in]{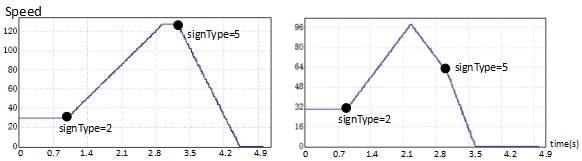}
  \caption{The simulation result of stop}
  \label{stopsmc}
\end{figure}

In Fig.\ref{stopsimu}, R7 and R9 can be validated. No matter whether the vehicle is driving at a constant speed, or the vehicle is decreasing its speed, if a \emph{stop} sign is detected, the vehicle will decrease its speed and stop at last. Same requirements can also be valid from Fig.\ref{stopsmc}, as whenever the vehicle recognises a stop sign, the vehicle will decrease its speed to stop.



\begin{figure}[ht!]
  \centering
  \includegraphics[width=4in]{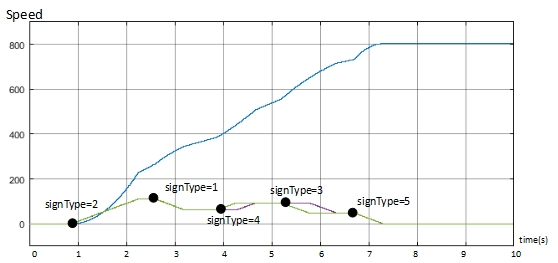}
  \caption{The simulation result of {\gt{energy}} consumption}
  \label{fig:ec}
\end{figure}

\begin{figure}[ht!]
  \centering
  \includegraphics[width=4in]{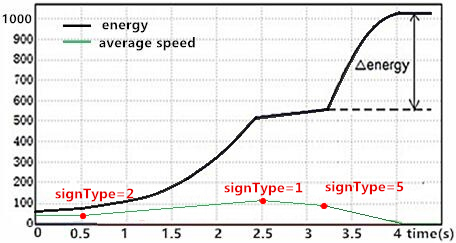}
  \caption{Energy Consumption Estimation in \smc}
  \label{esen}
\end{figure}

How the energy consumption rate of {\gt{Controller}} changes and how the estimated energy consumption is influenced by the change are monitored using \simu\\ (Fig.\ref{fig:ec}) and \smc\ (Fig.\ref{esen}). In Fig.\ref{fig:ec}, \av\ detects the minimum speed limit sign (\emph{signType} = 2) and it increases its speed (green and red curve). Since energy rate is proportional to speed, the energy consumption (blue curve) is continuously raised until \av\ detects the maximum speed limit sign (\emph{signType} = 1). \av\ slows down after it reaches the maximum speed at 3.2s leading to a slow decrease in energy consumption accordingly. Then a turn left sign is recognised, thus the vehicle increases the speed of its right wheels, and then turn right due to a turn right sign. Energy consumption increases faster than in constSpeed mode. Then a stop sign is recognised at 6.6s, the energy consumption of the vehicle increases with a rapid increase rate. The energy consumption for each movement of the vehicle are: 223J (acceleration), 138J (deceleration), 125J (left turn), 150J (right turn) and 80J (stop). Thus R40-R45 can be validated. In Fig.\ref{esen}, same behaviour of the \av~can be seen, and according energy consumption can be proved.

\begin{figure}
  \centering
  \subfigure[Simulation in Simulink]{\label{fig:ES}
  \includegraphics[width=4in]{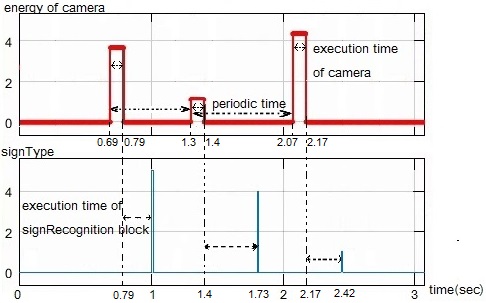}}
  \subfigure[Simulation in \smc]{
  \label{fig:EU}
  \includegraphics[width=4.3in]{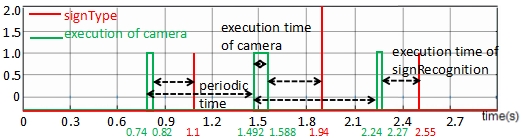}}
  \caption{Simulation of {\gt{Execution}} and {\gt{Periodic}} Constraints}
  \label{fig:et}
\end{figure}

Fig.\ref{fig:et} illustrates the simulation results of {\gt{execution}} constraints (R46, R47) in \simu\ and in \smc: In the upper graph (Fig.\ref{fig:et}.(a)), {\gt{Camera}} consumes energy within 100ms (red line), which demonstrates R46. is valid. In the lower graph, the time between the occurrences whereby {\gt{Sign Recognition}} receives a detected image from {\gt{Camera}} and sends out its corresponding sign type (blue line) to {\gt{Controller}} is within [200ms,600ms] thus validating R46. Similarly, in the simulation of \smc\ (Fig.\ref{fig:et}.(b)), the execution times of {\gt{Camera}} (green line) and {\gt{Sign Recognition}} (distance between green and red lines) satisfy R47 and R46 respectively.

Furthermore, \begin{inparaenum} \item {\gt{Camera}} captures an image almost every 700ms with a deviation of 100ms; \item The time intervals among different images {\gt{Camera}} captured are within 700ms with the deviation 100ms \end{inparaenum}. Both are illustrated in Fig.\ref{fig:et}.(a). Therefore, {\gt{Periodic}} constraint (R49) is established as valid.

\begin{figure}
  \centering
  \subfigure[Simulation in Simulink]{\label{fig:syS}
  \includegraphics[width=4.3in]{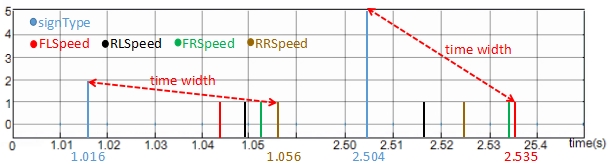}}
  \subfigure[Simulation in \smc]{
  \label{fig:syU}
  \includegraphics[width=4.3in]{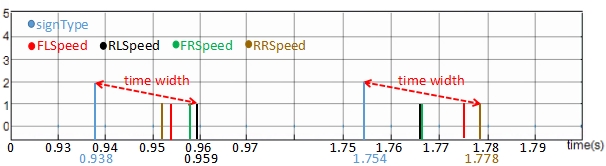}}
  \caption{Simulation of {\gt{Synchronization}} Constraint}
  \label{fig:synct}
\end{figure}

Fig.\ref{fig:syS} demonstrates {\gt{Synchronization}} constraint (R48) is valid: The blue line denotes the time point of which {\gt{Controller}} detects a sign type. The other lines (represented by four different colors) indicate the time points of which {\gt{Controller}} recognizes the speed signals of the four wheels. In both Fig.\ref{fig:syS}.(a) and (b), the earliest event arrival time and the lastest event arrival time among all input events are within the 400ms tolerance.

\begin{figure}
  \centering
  \includegraphics[width=4.5in]{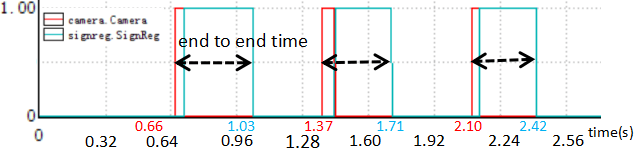}
  \caption{Simulation of { \gt {End-to-End}} constraint in \smc}
  \label{fig_e2esim}
\end{figure}

Fig.\ref{fig_e2esim} illustrates the simulation result of {\gt{end-to-end}} constraint (R50) in \smc: The red (blue) line indicates whether {\gt{Camera}} ({\gt{SignRecognition}}) is executed or not. The time between  {\gt{Sign Recognition}} receives a detected image from {\gt{Camera}} (red line) and sends out its corresponding sign type (blue line) to {\gt{Controller}} is within [200ms, 600ms] thus validating R50. Recall Fig.\ref{fig:et}.(a), R50 is similarly valid in \simu.

\chapter{Related work}
\label{sec:r-work}

In the context of \ed, an effort on the integration of \ed\ and formal techniques based on timing constraints was investigated in several works \cite{qureshi2011,kress13,ksafecomp11}, which are however, limited to the executional aspects of system functions without addressing energy-aware behaviors. An earlier study \cite{kicca13,kiceccs13} was performed towards the formal analysis of energy-aware \ed\ models based on informal semantics of \ed\ architectural models. Kang \cite{ksac14}, Goknil et al. \cite{Goknil2013} and Seceleanu et al. \cite{seceleanu2017analyzing} defined the execution semantics of both the controller and the environment of industrial systems in \ccsl\ \cite{abs-ccsl} which are also given as mapping to \uppaal\ models amenable to model checking. In contrast to our current work, those approaches lack precise stochastic annotations specifying continuous dynamics in particular regarding different energy consumption rates during execution. Though, Kang et al. \cite{kiciea16,kapsec15} and Marinescu et al. \cite{Marinescu3762} present both simulation and model checking approaches of \simu\ and \smc\ on \ed\ models, neither formal specification nor verification of extended \ed\ timing constraints with probability were conducted. Our approach is a first application on the integration of \ed\ and formal V\&V techniques based on a composition of energy- and probabilistic extension of \ed/\tdl\ constraints.

There are several works that discuss formal analysis approaches to verify the requirements of the system model. Integrated in the Simulink environment, SDV is a powerful tool for verifying the property of Simulink/Stateflow model. In \cite{ali2012applying}, Ali etc. compared \smc~and SDV for usability and adaptability for the functional requirements. In our work, applicability of SDV is evaluated by adding a set of non-functional properties. We translate Simulink/Stateflow model to \smc~and make advantages of \smc~ to verify the non-functional properties, which to some extent helps make up the limitation for expressing non-functional properties of SDV.
Leitner\cite{sdv2spin} compares SDV with SPIN and concludes that SDV can only prove the properties written as assertions, but SPIN can also check the temporal properties. In our work, we found that not only assertions can be checked in SDV of MATLAB2014b, but also functionally temporal property can be verified. What's more, we use \smc~instead of SPIN as there are some stochastic events in our plant, and we verify probabilistic timing constraints of our system to reduce the consumed time of verification. Filipovikj et al. \cite{filipovikj2016simulink} focus on the verification of Simulink blocks and proposed a method to transform Simulink blocks to Uppaal SMC model, which is limited in tackling timing requirements. In contrast to our work, we translate Simlink/Stateflow model and propose method for timing constraints analysis in Uppaal SMC. Moreover, simulation in both Simulink/Stateflow model and Uppaal model are conducted to ensure the consistency of the translation.
\chapter{Conclusion}
\label{sec:conclusion}

We present an approach to perform non-functional properties verification and support stochastic analysis of an autonomous vehicle system at the early design phase:  \begin{inparaenum} \item \ed/\tdl2\ are used for structural, timing- and \fp's causality constraints; the execution semantics of timing- and energy constraints are visualized in \simu\ and specified in \smc; \item probabilistic extension of \ed\ constraints is defined and the semantics of the extended constraints is translated into verifiable \smc\ models with stochastic semantics for formal verification; \item A set of mapping rules is proposed to facilitate the guarantee of translation. Simulation and V\&V are performed on the extended timing and energy constraints using \smc\ and \simu; \item The applicability of our approach in an autonomous automotive product is demonstrated. \end{inparaenum}

We discuss open issues: \begin{inparaenum}\item Although we have shown that our approach preserves ERT behaviors and the probabilistic extension of timing constraints of the \av\ model and that the obtained \smc\ and \simu\ models manifest the same behaviors and constraints, there is no formal correctness proof for the derived procedures. As ongoing work, we use conformance checking to show that the translation procedure correctly preserves probabilistic ERT behaviors and constraints; \item From the tooling perspective, a dedicated plugin that directly provides translation of \simu/\staf\ to \smc\ in fully automatic will supplement our tool chain, A-BeTA (A$\beta$: EAST-\textbf{A}DL \textbf{B}ehavioral Modeling and \textbf{T}ranslation into \textbf{A}analyzable Model) \cite{kiceccs13,ksac14,kqsic12} for both model checking and simulation. We formally prove the validity of the obtained \smc\ \end{inparaenum}

\chapter*{Acknowledgment}
This work is supported by the National Natural Science Foundation of China and International Cooperation \& Exchange Program (46000-41030005) within the project EASY.

\end{document}